\documentclass{ws-ijmpd}
\usepackage[super,compress]{cite}
\usepackage[dvips,breaklinks]{hyperref}
\hypersetup{colorlinks,urlcolor=black,citecolor=black,linkcolor=black,filecolor=black}

\usepackage{breakurl}
\usepackage{amsmath}
\usepackage{amsfonts}
\usepackage{amssymb}
\usepackage{graphicx}

\begin{document}

\markboth{Yu.L. Bolotin et al.}
{Applied (Practical) Cosmography}

%
\catchline{}{}{}{}{}
%

\title{APPLIED COSMOGRAPHY: A Pedagogical Review}

\author{Yu. L.  BOLOTIN}

\address{NSC ''Kharkov Institute of Physics and Technology'', Akademicheskaya Str. 1\\
Kharkiv, 61108,
Ukraine and \\
V.N.Karazin Kharkiv National University, 61022 Kharkov, Ukraine\\
ybolotin@gmail.com}

\author{V. A. Cherkaskiy}

\address{NSC ''Kharkov Institute of Physics and Technology'', Akademicheskaya Str. 1\\Kharkiv, 61108,Ukraine\\
vcherkaskiy@gmail.com}

\author{O.Yu. IVASHTENKO}

\address{V.N.Karazin Kharkiv National University, 61022 Kharkov, Ukraine\\
arinaivashtenko@gmail.com}

\author{M. I.  KONCHATNYI}
\address{NSC ''Kharkov Institute of Physics and Technology'', Akademicheskaya Str. 1\\
Kharkiv, 61108,
Ukraine and \\
V.N.Karazin Kharkiv National University, 61022 Kharkov, Ukraine\\
konchatnij@kipt.kharkov.ua}
\author{L. G. ZAZUNOV}

\address{NSC ''Kharkov Institute of Physics and Technology'', Akademicheskaya Str. 1\\Kharkiv, 61108,Ukraine\\
lzazunov@gmail.com}

\maketitle

\begin{history}
\received{Day Month Year}
\revised{Day Month Year}
\end{history}

\begin{abstract}
Based on the cosmological principle only, the method of describing the evolution of the Universe, called cosmography, is in fact a kinematics of cosmological expansion. The effectiveness of cosmography lies in the fact that it allows, based on the results of observations, to perform a rigid selection of models that do not contradict the cosmological principle. It is important that the introduction of new components (dark matter, dark energy or even more mysterious entities) will not affect the relationship between the kinematic characteristics (cosmographic parameters)
This paper shows that within the framework of cosmography the parameters of any model that satisfies the cosmological principle (the universe is homogeneous and isotropic on large scale), can be expressed through cosmographic parameters. The proposed approach to finding the parameters of cosmological models has many advantages. Emphasize that  all the  obtained results  are accurate, since they  follow from  identical transformations. The procedure can be generalized to the case of models with interaction between components

\end{abstract}


\ccode{PACS numbers:98.80.−k, 95.36.+x}

\tableofcontents
\section{Introduction}
Our current understanding of the structure of the Universe is based on the two fundamental models: the Standard Model of elementary particles\cite{Okun:1982ap,yndurain2014quantum} and the Standard Cosmological Model (SCM)\cite{Dodelson:1282338,weinberg2008cosmology,Mukhanov:991646,Gorbunov:2011zzc}. The first model is a combination of quantum chromodynamics with electroweak theory. Formulated in the 1970s, by now it has been confirmed by a huge number of experiments. The recent discovery of the missing element---the Higgs boson \cite{Higgs:1964pj,Chatrchyan:2012xdj}---was a convincing proof of the validity of the Standard Model of elementary particles. However, these successes did not stop attempts to discover physics outside of the Standard Model, and such physics was discovered\cite{Riess:1998cb,Perlmutter:1998np} on completely different spatial scales. The observed dynamics of the Universe cannot be described with the help of particles of the Standard Model. The new entities---dark energy \cite{amendola2010dark,Copeland:2006wr} and dark matter \cite{Bertone:2010zza,2014dmcw.bookE,Bertone:2016nfn} ---are the main components of the SCM, which managed to solve the main problems of the previous cosmological model (the Big Bang model) and to describe a giant array of observations.

However, the staggering success of SCM, for which it even got the name \emph{Concordance Model of Cosmology, or Cosmic Concordance}, should not be misleading. The cosmological model stating that 95\% of our world consists of essences of unknown nature (70\% of dark energy and 25\% of dark matter) can only be an intermediate step on the way to understanding the structure of the Universe. The lack of information on the nature of the main components of the energy budget of the Universe, on the one hand, significantly expands the number of possible options for describing cosmological evolution, and on the other hand, makes one think about the mechanism for selecting the most appropriate models. Such a mechanism may turn out to be the Cosmological Principle, stating that on scales exceeding hundreds of megaparsecs, the Universe is homogeneous and isotropic, which allows choosing from the whole conceivable variety of models of the description of the Universe a narrow class of homogeneous and isotropic models. The Cosmological Principle enables us to construct the metric of the Universe and to take the first steps to the interpretation of the cosmological observations. The method to describe the evolution of the Universe based only on the Cosmological Principle is called the cosmography, and in fact, it represents kinematics of the cosmological expansion. The effectiveness of cosmography relies on the fact that it allows, based on the results of observations, to perform a rigorous selection of models that do not contradict the Cosmological Principle. It is important that the introduction of the new components (dark matter, dark energy or even more mysterious entities) will not affect the relationship between the kinematic characteristics (the cosmographic parameters).

The fundamental characteristics of the evolution of the Universe can be either kinematic (if they are extracted directly from the space-time metric) or dynamic (if they depend on properties of the fields that fill the Universe). The dynamic characteristics, of course, are model dependent, while the kinematic characteristics are more universal. In addition, the latter are spared from the uncertainties that arise when measuring physical quantities, such as, for example, energy densities. That is the reason why the kinematic characteristics are convenient for the description of the current expansion of the Universe.

The evolution of a homogeneous and isotropic Universe is described by the scale factor $a(t)$ which connects the Lagrangian (or comoving) coordinates $r$ with the physical Euler coordinates $R(t)$
\begin{equation}\label{eq1_2}
R(t)=a(t)r
\end{equation}
Differentiating this relation with respect to  time, we obtain the Hubble law
\begin{equation}\label{eq1_3}
V=HR
\end{equation}
where $V=dR/dt=\dot R$ is the speed of cosmological expansion and $H=\dot a/a$  is the Hubble parameter, relating the distance to the extragalactic object with the speed of its recession.

The expansion rate of the Universe, as determined by the Hubble parameter, generally speaking, depends on time. The measure of this dependence is the deceleration parameter. We define it using the Taylor series expansion of the scale factor in the neighbourhood of the current time moment $t_0$:
\begin{equation}\label{eq1_4}
a(t) = a(t_0)+\dot a(t_0)[t-t_0]+\frac12\ddot a(t_0)[t-t_0]^2.
\end{equation}
Let us represent this relation in the form
\begin{equation}\label{eq1_5}
\frac{a(t)}{a(t_0)}=1+H_0[t-t_0]-\frac{q_0}2H_0^2[t-t_0]^2.
\end{equation}
where the deceleration parameter reads
\begin{equation}\label{eq1_6}
q(t)\equiv-\frac{\ddot a(t)a(t)}{\dot a^2}=-\frac{\ddot a}{a(t)}\frac1{H^2(t)}.
\end{equation}
If we are interested solely in the expansion regime then in the case of constant deceleration parameter the Universe would exhibit decelerating expansion if $q > 0$, an expansion with constant rate if $q = 0$, accelerating power-law expansion if $−1 < q < 0$, exponential expansion (also known as the de Sitter expansion) if $q = −1$ and super-exponential expansion if $q < −1$.
In the early seventies, Alan Sandage \cite{1962ApJSandage} defined cosmology as the search for two numbers: the Hubble parameter $H_0$ and the deceleration parameter $q_0$. It seemed too simple and clear: the main term in form of the Hubble parameter determined the expansion rate of the Universe and a small correction due to the gravity of the matter content was responsible for slow down of the expansion. However, the situation drastically changed at the end of the last century.

Expansion with a constant acceleration is the simplest, but not the only possible realization of the kinematics of a non-stationary Universe. As the Universe evolves, relative contents of its components change, which affect dynamics of the expansion, and as a result the acceleration value varies. In order to track this variation, it is necessary to consider the third derivative of the scale factor with respect to time. Obviously, the inevitable progress in the accuracy of cosmological observations will require the inclusion of higher time derivatives of the scale factor into consideration. In other words, instead of the simplest decomposition (\ref{eq1_4})
we must take into account the terms of the Taylor series for the scale factor of higher order ($N>2$)
\begin{equation}\label{eq1_7}
a(t)=\lim\limits_{N\to\infty}\sum\limits_{n=0}^N\frac{a^{(n)}}{n!}(t-t_0)^n.
\end{equation}
The present work uses the scheme of the description of the Universe, called the ''cosmography'' \cite{Weinberg:1972kfs}, entirely based on the Cosmological Principle. Recall that in the Classical Mechanics, the kinematics is understood as the part that describes the motion of bodies, irrespective of the forces that could cause it. In this sense, the cosmography merely represents kinematics of the cosmological expansion. In fact, the term "cosmo-kinematics" is a synonym to the term "cosmography". Inclusion of higher-order ($N>2$) terms in the Taylor series (\ref{eq1_6}) allows us to go beyond the frames of the motion with a constant acceleration, and to transform the cosmography into a cosmic kinematics of general case.

To construct the key cosmological characteristic---the time dependence of the scale factor $a(t)$---we need the equations of motion (the Einstein's equations), and the assumption on the material composition of the Universe, which enables us to construct the energy-momentum tensor. The cosmography provides an effective and universal way to compare the model solutions (like the solutions of General Relativity) with the observations. The latter provides us with a set of cosmological parameters, which must be compared with the values of the parameters calculated within the framework of a certain model. The result of the comparison is a conclusion about the adequacy of the corresponding models.

Thus, for a more complete description of kinematics of the cosmological expansion, it is useful to consider the extended set of parameters, including the higher-order time derivatives of the scale factor \cite{Visser:2004bf,Visser:2003vq,Dunsby:2015ers,Bolotin:2012yja}
\begin{align}
\nonumber H(t)&\equiv\frac1a\frac{da}{dt};\\
\nonumber q(t)&\equiv-\frac1a\frac{d^2a}{dt^2}\left[\frac1a\frac{da}{dt}\right]^{-2};\\
\label{eq1_8} j(t)&\equiv\frac1a\frac{d^3a}{dt^3}\left[\frac1a\frac{da}{dt}\right]^{-3};\\
\nonumber s(t)&\equiv\frac1a\frac{d^4a}{dt^4}\left[\frac1a\frac{da}{dt}\right]^{-4};\\
\nonumber l(t)&\equiv\frac1a\frac{d^5a}{dt^5}\left[\frac1a\frac{da}{dt}\right]^{-5};
\end{align}
Note that the last four parameters are dimensionless. It should be noted that the jerk parameter $j(t)$ is of unique importance (especially for testing cosmological models). ''It is a striking and slightly puzzling fact that almost all current cosmological observations can be summarized by the simple statement: The jerk of the universe equals one''\cite{Dunajski:2008tg}. This parameter determines evolution of the deceleration parameter. Inclusion of the higher order derivatives of the scale factor, on the one hand, reflects the continuous progress of the observational cosmology, and, on the other, is dictated by the need to describe the increasingly complex effects used to obtain the precise observational data.

In terms of the cosmographic parameters introduced above (\ref{eq1_7}), the decomposition (\ref{eq1_6}) (up to terms of the fifth order) takes on the form
\begin{align}\label{eq1_9}
\nonumber a(t)&=a_0\left[1+H_0[t-t_0]-\frac12q_0H_0^2[t-t_0]^2+\frac1{3!}j_0H_0^3(t-t_0)^3\right.\\
&\left.+\frac1{4!}s_0H_0^4(t-t_0)^4+\frac1{5!}l_0H_0^5(t-t_0)^5+O\left((t-t_0)^6\right)\right],
\end{align}
where $a_0$ is the present value of the scale factor, and $z$ is the redshift.

Note that a simplified version of the cosmographic parameter set is represented by the so-called statefinder parameters $\{r,s\}$ introduced by Sahni, Saini, and Starobinsky \cite{Sahni:2002fz}:
\begin{equation}\label{eq1_10}
r=\frac{\dddot a}{aH^3},\quad s\equiv\frac{r-1}{3(q-1/2)}
\end{equation}
The statefinders are dimensionless and similarly to the cosmographic parameters they are constructed from the scale factor and its time derivatives. The statefinder $r$ exactly coincides with the cosmographic jerk parameter $j$, and $s$ is a combination of the parameters $q$ and $r$ (or $j$). The combination is chosen in such a way that it does not depend on the density. The values of the statefinders can be reconstructed from the available cosmological data. After that, the statefinders can be successfully used to identify various models of dark energy. 
\section{\label{alphabet}Cosmographic alphabet: relations between the cosmographic parameters}
The parameter set $H,q,j,s,l\ldots$ represents the alphabet of the cosmography. Naturally the next step should be to use the alphabet in order try building words and sensible phrases based on certain grammar rules. Relations between the cosmographic parameters stand for the grammatic rules in the cosmography.

Let us use the definition (\ref{eq1_6}) of in order to establish some useful relation between the deceleration parameter and the Hubble parameter\cite{Dunsby:2015ers,Bolotin:2012yja}:
\begin{align}
\nonumber q(t)&=\frac{d}{dt}\left(\frac{1}{H}\right)-1,\\
\nonumber q(z)&=\frac{1+z}{H}\frac{dH}{dz}-1,\\
\nonumber q(z)&=\frac12(1+z)\frac{1}{H^2}\frac{dH^2}{dz}-1,\\
\nonumber q(z)&=\frac12\frac{d\ln H^2}{d\ln(1+z)}-1\\
\label{alph.eq1} q(z)&=(1+z)\frac{d\ln H}{d\ln z}-1,\\
\nonumber q(a)&=-\left(1+\frac{\frac{dH}{dt}}{H^2}\right),\\
\nonumber
q(a)&=-\left(1+\frac{a\frac{dH}{da}}{H}\right),\\
\nonumber
q(a)&=-\frac{d\ln(aH)}{d\ln a}.
\end{align}
The derivatives $dH/dz$, $d^2H/dz^2$, $d^3H/dz^3$, and $d^4H/dz^4$ can be expressed in terms of the deceleration parameter $q$ and other cosmographic parameters:
\begin{align}
\nonumber \frac{dH}{dz}&=\frac{1+q}{1+z}H;\\
\nonumber \frac{d^2H}{dz^2}&=\frac{j-q^2}{(1+z)^2}H;\\
\label{hubble_parameter_z_derivatives}\frac{d^3H}{dz^3}&=\frac{3q^2+3q^3-4qj-3j-3s}{(1+z)^3}H;\\
\nonumber \frac{d^4H}{dz^4}&=\frac{-12q^2-24q^3-15q^4+32qj+25q^2j+7qs+12j-4j^2+8s+l}{(1+z)^4}H.
\end{align}
The expresions (\ref{hubble_parameter_z_derivatives}) allow to decompose the Hubble parameter in the Taylor series over the redshift:
\begin{align}
\label{alph.eq2} H(z)&=H_0+\left.\frac{dH}{dz}\right|_{z=0}z+\left.\frac12\frac{d^2H}{dz^2}\right|_{z=0}z^2+\left.\frac16\frac{d^3H}{dz^2}\right|_{z=0}z^3+\dots\\
\nonumber =H_0&\left(1+(1+q_0)z+\frac{z^3}2(j_0-q_0^2)+\frac{z^3}6(3q_0^2+3q_0^3-4q_0j_0-3j_0-3s_0)\right)
\end{align}
The inverse of the Hubble parameter can be decomposed as well:
\begin{align}
\nonumber \frac{d}{dz}\left(\frac1H\right)&=-\frac1{H^2}\frac{dH}{dz}=-\frac{1+q}{1+z}\frac1H;\\
\label{alph.eq3} \frac{d^2}{dz^2}\left(\frac1H\right)&=2\left(\frac{1+q}{1+z}\right)^2\frac1H-\frac{j-q^2}{(1+z)^2}\frac1H=\frac{2+4q+3q^2-j}{(1+z)^2}\frac1H;\\
\nonumber \frac1{H(z)}&=\frac1{H_0}\left[1-(1+q_0)z+\frac{z^2}6(2+4q_0+3q_0^2-j_0)+\dots\right].
\end{align}
The following relations are useful to transit from higher order time derivatives to the derivatives with respect to redshift:
\begin{align}
\nonumber \frac{d^2}{dt^2}&=(1+z)H\left[H+(1+z)\frac{dH}{dz}\right]\frac{d}{dz}+(1+z)^2H^2;\\
\nonumber \frac{d^3}{dt^3}&=-(1+z)H\left\{H^2+(1+z)^2\left(\frac{dH}{dz}\right)^2+(1+z)H\left[4\frac{dH}{dz}+(1+z)\frac{d^2H}{dz^2}\right]\right\}
\frac{d}{dz}\\
\nonumber &-3(1+z)^2H^2\left[H+(1+z)\frac{dH}{dz}\right]\frac{d^2}{dz^2}-(1+z)^3H^3\frac{d^3}{dz^3};\\
\label{alph.eq4}
\frac{d^4}{dt^4}&=(1+z)H\left[H^2+11(1+z)H^2\frac{dH}{dz}+11(1+z)H\frac{dH}{dz}+(1+z)^3\left(\frac{dH}{dz}\right)^3\right.\\
\nonumber &\left.+7(1+z)^2H\frac{d^2H}{dz^2}+4(1+z)^3H\frac{dH}{dz}\frac{d^2H}{dz^2}+(1+z)^3H^2\frac{d^3H}{dz^3}\right]\frac{d}{dz}\\
\nonumber
&+(1+z)^2H^2\left[7H^2+22H\frac{dH}{dz}+7(1+z)^2\left(\frac{dH}{dz}\right)^2+4H\frac{d^2H}{dz^2}\right]\frac{d^2}{dz^2}\\
\nonumber
&+6(1+z)^3H^3\left[H+(1+z)\frac{dH}{dz}\right]\frac{d^3}{dz^3} +(1+z)^4H^4\frac{d^4}{dz^4} +(1+z)^4H^4\frac{d^4}{dz^4}.
\end{align}
Derivatives of the Hubble parameter squared with respect to the redshift \[\frac{d^iH^2}{dz^{i}},\ i=1,2,3,4\] can be expressed through the cosmographic parameters take on the form:
\begin{align}
\nonumber \frac{d(H^2)}{dz}&=\frac{2H^2}{1+z}(1+q);\\
\label{alph.eq5}\frac{d^2(H^2)}{dz^2}&=\frac{2H^2}{(1+z)^2}(1+2q+j);\\
\nonumber \frac{d^3(H^2)}{dz^3}&=\frac{2H^2}{(1+z)^3}(-qj-s);\\
\nonumber \frac{d^4(H^2)}{dz^4}&=\frac{2H^2}{(1+z)^4}(4qj+3qs+3q^2j-j^2+4s+l).
\end{align}
Current values of the parameters $q$ and $j$ in terms of the derivatives with respect to $N=-\ln(1+z)$ are
\begin{align}
\nonumber q_0&=-\frac1{H^2}\left.\left\{\frac12\frac{d(H^2)}{dN}+H^2\right\}\right|_{N=0},\\
\label{alph.eq6} j_0&=\left.\left\{\frac1{2H^2}\frac{d^2(H^2)}{dN^2}+\frac3{2H^2}\frac12\frac{d(H^2)}{dN}+1\right\}\right|_{N=0}.
\end{align}
It is important for further consideration to have the expressions of the time derivatives of the Hubble parameter in terms of the cosmographic parameters:
\begin{align}
\nonumber\dot H&=-H^2(1+q);\\
\label{hubble_paramters_time_derivatives}\ddot H&=H^3(j+3q+2);\\
\nonumber\dddot H&=H^4(s-4j-3q(q+4)-6);\\
\nonumber\ddddot H&=H^5(l-5s+10(q+2)j+30(q+2)q+24).
\end{align}
Let \[C_n\equiv\gamma_n\frac{a^{(n)}}{aH^n},\] where $a^{(n)}$ is $n$-th derivative of the scale factor with respect to time, $n\ge2$ and $\gamma_2=-1$, $\gamma_n=1$ for $n\ne2$. Then $C_2=q$, $C_3=j$, $C_4=s\ldots$. Then derivatives of the parameters $C_n$ with respect to the redshift satisfy the relation:
\[(1+z)\frac{dC_n}{dz}=-\frac{\gamma_n}{\gamma_{n+1}}C_{n+1}+C_n-nC_n(1+q).\]
Combining the above mentioned recurrent relation with the formula \[\frac{d}{dt}=-H(1+z)\frac{d}{dz},\]
one can obtain the time derivatives of the cosmographic parameters in th form
\begin{align}
\nonumber\frac{dq}{dt}&=-H(j-2q^2-q);\\
\label{alph.eq8}\frac{dj}{dt}&=H(s+j(2+3q));\\
\nonumber\frac{ds}{dt}&=H(l+s(3+4q));\\
\nonumber\frac{dl}{dt}&=H(m+l(4+5q)).
\end{align}
This relation can be used to express higher order cosmographic parameters in terms of lower ones and its derivatives:
\begin{align}
\nonumber j&= -q+2q(1+q)+(1+z)\frac{dq}{dz};\\
\label{alph.eq9} s&= j-3j(1+q)-(1+z)\frac{dj}{dz};\\
\nonumber l&= s-4s(1+q)-(1+z)\frac{ds}{dz};\\
\nonumber m&= l-5l(1+q)-(1+z)\frac{dl}{dz}.
\end{align}
Combining the expressions for time derivatives of the Hubble parameter with the Friedmann equations, we obtain density and pressure of one-component flat Universe in terms of the cosmographic parameters:
\begin{align}
\nonumber \rho&=3H^2;\\
\nonumber \frac{d\rho}{dt}&=-6H^3(1+q);\\
\nonumber \frac{d^2\rho}{dt^2}&=6H^4(j+q(q+5)+3);\\
\label{alph.eq10} \frac{d^3\rho}{dt^3}&=6H^5[s-j(3q+7)-3q(4q+9)-12];\\
\nonumber p&=-H^2(1-2q);\\
\nonumber \frac{dp}{dt}&=-2H^3(j-1);\\
\nonumber \frac{d^2p}{dt^2}&=-2H^4(s-j+3q+3);\\
\nonumber \frac{d^3p}{dt^3}&=-2H^5[l-j(q+1)-3q(2q+7)-2(6+s)].
\end{align}
As an example, let derive expression for $dp/dt$. Using the second Friedmann equation, one finds:
\begin{align}
\nonumber \dot H +H^2&=\frac{\ddot a}{a}=-\frac16 (\rho+3p)=-\frac12 H^2-\frac12 p;\\
\label{alph.eq11} \ddot H+3H\dot H&=-\frac12 \dot p.
\end{align}
Using the expressions for $\dot H$ and $\ddot H$ from (\ref{hubble_paramters_time_derivatives}), one obtains:
\[\frac{dp}{dt}=-2H^3(j-1).\]
It is possible also to obtain expression in terms of cosmographic parameters for the scalar curvature $R=g^{\mu\nu}R_{\mu\nu}$ and its time derivatives. Combining the expression \[R=-6\left(\frac{\ddot a}{a}+H^2\right)\] with the definition (\ref{eq1_6}) of the deceleration parameter, one obtains:
\[R=-6H^2(1-q).\]
Taking time derivative of the above expression and using (\ref{hubble_paramters_time_derivatives}), obtains the expressions for the time derivatives of the scalar curvature:
\begin{align}
\nonumber \dot R&=-6H^3(j-q-2);\\
\label{alph.eq12} \ddot R&=-6H^4(s+q^2+8q+6).
\end{align}
The relations between the cosmographic parameters obtained above open new possibilities to analyze evolution of the Universe and to test a wide class of cosmological models. 
\section{Cosmographic parameters for some cosmological models}
Let us briefly consider the cosmography of a number of the simplest cosmological models. Of course, we focus on exclusively kinematic features of the models.
\subsection{Models with the power-law time dependence of the scale factor}
Let us start from the widely used model with the power-law time dependence of the scale factor (the power-law cosmology) \cite{Kumar:2011sw,Rani:2014sia,Dolgov:2014faa}
\begin{equation}
\label{eq1_100}
a(t)=a_0\left(\frac{t}{t_0}\right)^\alpha,
\end{equation}
where $\alpha$ is a dimensionless positive parameter. The deceleration parameter thus equals
\begin{equation}
\label{eq1_101}
q(t)\equiv-\frac{\ddot a}{aH^2}=\frac1\alpha-1.
\end{equation}
The considered model describes a uniformly accelerated expansion ($q=\mathrm{const}$) under the condition that $\alpha>0$ (which transforms into $q>-1$). The history of the expansion of the Universe $H(z)$ is described in this model by the two parameters ($H_0, q$):
\begin{equation}
\label{eq1_102}
H(z)=H_0(1+z)^{1+q}.
\end{equation}
In the case of a power-law of expansion for $\alpha<1$ (decelerated expansion), the Hubble radius grows faster than the scale factor of the expansion of the Universe: $R_H=H^{-1}\propto t$, while $a(t)\propto t^\alpha$. Just such a situation, as we saw above, takes place for matter ($\alpha=2/3$) and radiation ($\alpha=1/2$).

\subsection{Hybrid expansion law}

We now consider a simple generalization of the model with a power law of expansion, called the hybrid expansion law \cite{Akarsu:2013xha}
\begin{equation}\label{eq1_103}
a(t)=a_0\left(\frac{t}{t_0}\right)^\alpha e^{\beta\left(\frac{t}{t_0}-1\right)},
\end{equation}
where $\alpha$ and $\beta$ are nonnegative constants. Here $a_0$ and $t_0$, as in the case of the power-law of expansion, represent the current value of the scale factor and the age of the Universe. In special cases $\alpha=0$ and $\beta=0$, the hybrid expansion law reduces to exponential and power-law expansion, respectively.

Modification of equation (\ref{eq1_103}) allows us to go beyond the limits of uniformly accelerated expansion and leads to the time dependence of all cosmographic parameters. In particular, for the three lowest cosmological parameters we find
\begin{align}
\nonumber H&=\frac\alpha t+\frac\beta{t_0};\\
\label{eq1_104} q&=\frac{\alpha t_0^2}{(\beta t+\alpha t_0)^2}-1;\\
\nonumber j&=\frac{\alpha t_0^2(2t_0-3\beta t-3\alpha t_0)}{(\beta t+\alpha t_0)^3}+1.
\end{align}
Below we give the asymptotes of the scale factor and cosmological parameters for the hybrid expansion law at $t\to0$ and $t\to\infty$:
\begin{align}
\label{eq1_105}
t\to&0\Rightarrow a\to a_0\left(\frac{t}{t_0}\right)^\alpha,\quad H\to\frac\alpha t,\quad q\to-1+\frac1\alpha,\quad j\to1-\frac3\alpha+\frac2{\alpha^2};\\
\label{eq1_106}
t\to&\infty\Rightarrow a\to a_0e^{\beta\left(\frac{t}{t_0}-1\right)},\quad H\to\frac\beta{t_0},\quad q\to-1,\quad j\to1.
\end{align}
The time moment at which the hybrid model transits from the decelerated expansion to the accelerated one is determined from (\ref{eq1_104}):
\begin{equation}\label{eq1_107}
q=0\Rightarrow\frac{t_{tr}}{t_0}=\frac{\sqrt\alpha-\alpha}{\beta}.
\end{equation}

\subsection{Chaplygin gas}

Numerous attempts to explain the observed accelerated expansion of the Universe in the framework of General Relativity stimulated the consideration of the components with "non-traditional" equations of state. The component with the name "Chaplygin gas" (the name is borrowed from aerodynamics) realizes one of such possibilities \cite{Kamenshchik:2001cp}. The equation of state for this component reads
\begin{equation}\label{eq1_108}
p=-\frac{A}{\rho},
\end{equation}
where $A>0$ is a constant. The desire to improve the description of the observed dynamics of the Universe led to numerous generalizations of the equation of state (\ref{eq1_108}).

The equation of state for the so-called generalized Chaplygin gas \cite{Bento:2002ps,Bilic:2001cg} reads
\begin{equation}\label{eq1_109}
p=-\frac{A}{\rho^\alpha},
\end{equation}
where $0\le\alpha\le1$ (the additional parameter $\alpha$ extends the capabilities of the model). A further modification of the equation of state (\ref{eq1_108}) was connected with the inclusion of a positive term linear in density:
\begin{equation}\label{eq1_110}
p=B\rho-\frac{A}{\rho^\alpha}.
\end{equation}
Here $A$, $B$, and $\alpha$ are positive constants and $0\le\alpha\le1$. Such a model was called the modified Chaplygin gas \cite{Benaoum:2002zs}. An attractive feature of the cosmological models based on the equation of state (\ref{eq1_108}) and its generalizations is the ability to describe in a single equation the states of a substance with essentially different physical properties. For example, the modified Chaplygin gas at low density limit describes a substance generating a negative pressure (can be used to describe the current evolution of the Universe), while at high densities and $B=1/3$ or $B=0$ this state equation represents a description of radiation or non-relativistic matter with zero pressure (can be used to describe the early Universe).

Integrating the conservation equation, we find the density of the modified Chaplygin gas:
\begin{equation}\label{eq1_111}
\rho_{MCG}=\rho_0\left[A_S+\frac{1-A_S}{a^{3(1+B)(1+\alpha)}}\right]^\frac1{1+\alpha}.
\end{equation}
Here \[A_S\equiv\frac{A}{1+B}\frac{1}{\rho_0^{\alpha+1}},\quad B\ne1\] and $\rho_0$ is the constant of integration. In the considered model of the Universe it is required to introduce only the baryonic component (there is no need to introduce the dark matter), since the Chaplygin gas represents a unifying model of the dark components. 
Using the definition of the deceleration parameter, we find that for a flat Universe filled a single component with the equation of state $p = w\rho, $
\begin{equation}\label{eq1_111n1}
q = \frac{1}{2}(1 + 3w).
\end{equation}
In general, \[
(k = 0, \pm 1,{\kern 1pt} \quad \rho  = \sum\limits_i  \rho _i ,\quad p = \sum\limits_i  \rho _i w_i )
\]
 obtain
\begin{equation}\label{eq1_111n2}
q = \frac{\Omega }{2} + \frac{3}{2}\sum\limits_i  w_i \Omega _i .
\end{equation}
Here
	\[
\Omega _i  \equiv \frac{{\rho _i }}{{\rho _c }};{\kern 1pt} \quad \rho _c  \equiv 3M_{Pl}^2 H^2 \quad \Omega  \equiv \sum\limits_i  \Omega _i .\quad
\]

Using (\ref{eq1_111n2}), we find the deceleration parameter for a spatially flat universe filled with a modified Chaplygin gas:
\begin{align}
\nonumber
q(z)&=\frac{\frac{\Omega_{b0}}{a^3}+\Omega(z)[1+3w(z)]}{2\left[\frac{\Omega_{b0}}{a^3}+\Omega(z)\right]};\\
\label{eq1_112}
\Omega(z)&=\Omega_0\left[A_S+(1-A_S)(1+z)^{3(1+B)(1+\alpha)}\right]^\frac{1}{1+\alpha}.
\end{align}
Here $\Omega_b$ and $\Omega$ are relative densities of the baryonic component and the modified Chaplygin gas respectively.

The parameter of the equation of state of the modified Chaplygin gas can be found using the expression (\ref{eq1_110}):
\begin{equation}\label{eq1_113}
w\equiv\frac p\rho=B-\rho^{-(\alpha+1)}=B-\frac{A_S(1+B)}{A_S+(1-A_S)(1+z)^{3(1+B)(1+\alpha)}}.
\end{equation}
As for the deceleration parameter $q(z)$, it, in turn, is determined by the three parameters $A$, $B$ and $\alpha$ whose values can be calculated by analyzing traditional sets of the observational data. The values of the parameters obtained in this way \cite{Paul:2014kza} satisfy the relations: $A_S\approx0.8$, $B\le0.1$, and $\alpha\le0.2$. Substituting them into (\ref{eq1_112}) and (\ref{eq1_113}), we obtain the values of $w(z=0)=-0.8$ and $q(z=0)=-0.7$, which, like the dependence $q(z)$, are close to the values obtained in the SCM.

\subsection{Scalar fields}

The cosmological constant represents just one of the possible realizations of the hypothetical substance - the dark energy introduced to explain the accelerated expansion of the Universe. Unfortunately, the nature of dark energy is unknown to us, which generates a huge number of hypotheses and candidates for the role of a fundamental component of the energy budget of the Universe. We mentioned many times about the rapid progress of observational cosmology in the last decade. However, we still can not answer the question of the temporal evolution of the density of dark energy. If this value varies with time, we are forced to seek an alternative to the cosmological constant. In a very short time, many alternative possibilities have been explored. Scalar fields \cite{Gorbunov:2011zzc,amendola2010dark,Copeland:2006wr}, which formed the post-inflationary Universe, are one of the main candidates for the role of dark energy. The most important discovery in elementary particle physics---the discovery of the Higgs boson---has significantly strengthened the status of the scalar fields.

The most popular version is a scalar field $\varphi$ with a suitably selected potential $V(\varphi)$. In these models, unlike the cosmological constant, the scalar field is a dynamic variable, and the density of dark energy depends on time. The models differ in the choice of the Lagrangian of the scalar field. Let us start with, perhaps, the simplest model of the dark energy of this type, called the quintessence \cite{Ratra:1987rm, Wetterich:1987fm, Caldwell:1997ii,Zlatev:1998tr,Caldwell:2005tm}. By the quintessence we mean a scalar field $\varphi$ in a potential $V(\varphi)$ minimally coupled to gravity, i.e. it is only influenced by the curvature of space-time, and it is limited by the canonical form of kinetic energy. The action for such a field has the form
\begin{equation}\label{eq1_114}
S\equiv\int d^4x\sqrt{-g}L=\int d^4x\sqrt{-g}\left[\frac12g^{\mu\nu}\frac{\partial\varphi}{\partial x_\mu}\frac{\partial\varphi}{\partial x_\nu}-V(\varphi)\right],
\end{equation}
where $g\equiv\det g_{\mu\nu}$. The equation of motion for the scalar field is found by varying the action over the field,
\begin{equation}\label{eq1_115}
\frac1{\sqrt{-g}}\partial_\mu\left(\sqrt{-g}g^{\mu\nu}\frac{\partial\varphi}{\partial x_\nu}\right)=-\frac{dV}{d\varphi}.
\end{equation}
In the flat Friedmannian Universe, i.e. for the FLRW metric, for the homogeneous field $\varphi(t)$ we obtain the equation:
\begin{equation}\label{eq1_116}
\ddot\varphi+3H\dot\varphi+V'(\varphi)=0.
\end{equation}
In the case of the homogeneous field $\varphi(t)$ in a locally Lorentzian system in which the metric $g_{\mu\nu}$ can be replaced by the Minkowski metric, we obtain for the density and pressure of the scalar field
\begin{equation}\label{eq1_117}
\rho_\varphi\equiv T_{00}=\frac12\dot\varphi^2+V(\varphi);\quad p_\varphi\equiv T_{ii}=\frac12\dot\varphi^2-V(\varphi).
\end{equation}
The Friedmann equations for the flat Universe filled with the scalar field take on the form
\begin{align}
\nonumber
H^2&=\frac13\left[\frac12\dot\varphi^2+V(\varphi)\right];
\label{eq1_118}
\dot H=-\frac{\dot\varphi^2}{2}.
\end{align}
The deceleration parameter in the Universe filled with the scalar field is
\begin{equation}\label{eq1_119}
q_\varphi=\frac{d}{dt}\left(\frac1H\right)-1=\frac{\dot\varphi^2-V}{\frac12\dot\varphi^2+V}.
\end{equation}

\section{Cosmography as a universal method to analyze the cosmological models}
\subsection{Basic cosmographic methods for determination of the model parameters}

Dynamical system \cite{1997math} represents a mathematical model of some object, process or phenomenon, which enables us to describe its time evolution. However odd it could seem, an
introduction of some simplifying assumptions makes the Universe a rather simple dynamical system. In the framework of GR the dynamics of homogeneous and isotropic Universe obeying the cosmological principle can be reduced to the simple differential equations (the Friedmann equations) for the scale factor and the energy densities of its components. The Friedmann equations represent a mathematical model of the real Universe. A mathematical model in form of the dynamical system is considered well-defined if, besides the law of evolution for an initial state (the differential equations), it contains the parameters reflecting the parameters of the real object which we intend to model. The problem is that the model parameters are linked to the observed ones, but they are not identical. Numerous cosmological models used for description of evolution of the Universe generated even greater number "independent" parameters of different nature, such as initial density of matter and cosmological constant value in the SCM, equation of state parameters (in the polytropic model or Chaplygin gas), the coupling constant in different models with interacting dark energy and dark matter, matter creation rate (in the models with particle generation), dissipative characteristics (in the models with bulk viscosity), to mention a few. A natural question arises: is it possible to put this disordered mess into a system or reduce the number of the parameters if we establish the links between the parameters on a deeper level? It is due to recall here an important methodological principle, the so-called the Occam's razor: {\it there is no need to multiply essences without necessity}. This is more than an aesthetic question. Having reduced number of the cosmological parameters and developed a unified method of their determination, we will simplify testing of the models and limit the role of the observational cosmology to the determination of a small number of the cosmographic parameters.

In 2008 Dunaisky and Gibbons \cite{Dunajski:2008tg} proposed an original approach to testing of the cosmological models obeying the cosmological principle. The essence of the proposed approach is extremely simple: let the system is described by $n$ free parameters. Assuming that the dynamical variables are multiply differentiable functions of time, let us differentiate the initial evolution equation  $n$ times. The obtained system will be used to express the free parameters in terms of time derivatives of the dynamical variables. The latter can be treated as a set of kinematic parameters available to direct observation. An attractive aspect of this approach to the determination of the parameters is the necessity to solve a system of algebraic rather than differential equations. It should be stressed that thus obtained relations between the model parameters and the kinematic parameters are exact.

In order to realize the proposed approach treating the Universe as a dynamical system described by the Friedmann equations, the following steps should be made:
\begin{enumerate}
\item Transform the first Friedmann equation into an ordinary differential equation for the scale factor. In order to that we use the conservation equation for each component included into the model to determine the dependence of the energy density on the scale factor.
\item Differentiate the obtained equation as many times as there are free parameters in the model.
\item Express the time derivatives of the scale factor in terms of the cosmographic parameters.
\item Having solved the obtained system of linear algebraic equations, we express the free parameters of the model in terms of the cosmographic parameters.
\end{enumerate}
To give the realization examples of the considered procedure, let us start from the SCM---the model of the Universe filled by the dark energy in form of the cosmological constant and non-relativistic matter which do not interact with each other.

Using the expression for the matter density
\[\rho_m=\frac{M}{a^3}\quad (M=const),\]
we can represent the first Friedmann equation in the following form
\begin{equation}\label{eq2.1}
\dot{a}^2+k=\frac13\frac M a+\frac13\Lambda a^2, 8\pi G=1.
\end{equation}
Differentiation of the latter expression twice with respect to time gives
\begin{align}
\nonumber \ddot a =-\frac16\frac{M}{a^2}&+\frac13\Lambda a,\\
\dddot a =\frac13\frac{M\dot a}{a^3}&+\frac13\Lambda\dot a.
\label{eq2.2}
\end{align}
Using the definition of the cosmographic parameters
\begin{equation}
H\equiv\frac{\dot a}a,\quad q\equiv-a\frac{\ddot a}{\dot a^2},\quad j\equiv a^2\frac{\dddot a}{\dot a^3};
\label{eq2.3}
\end{equation}
the equation (\ref{eq2.2}) can be presented in the form
\begin{align}
\nonumber q&=\frac12A-B,\\
j&=A+B,\quad A \equiv\frac13\frac M{a^3H^2},\quad B\equiv\frac13\frac\Lambda{H^2}. \label{eq2.4}
\end{align}
One then finds that
\begin{align}
\nonumber A&=\frac23(j+q),\\
\label{eq2.5} B&=\frac23\left(\frac12 j-q\right).
\end{align}
The Friedmann equation (\ref{eq2.1}) in terms of the above introduced variables $A$ and $B$ takes on the following form:
\begin{equation}
\label{eq2.6}
\frac k{a^2}=(A+B-1)H^2,
\end{equation}
or equivalently
\begin{equation}
\label{eq2.7}
k=a^2H^2(j-1).
\end{equation}
The relation (\ref{eq2.7}) links the curvature parameter $k$ with the cosmographic parameters $H$ and $j$. It then follows that the condition $j=1$ is necessary to ensure that the Universe with the given energy composition has the flat geometry.

The expressions (\ref{eq2.4}) and (\ref{eq2.5}) solve the above posed question how to express the model parameters in terms of the cosmographic ones. In particular, for the energy density of the cosmological constant \[\rho_\Lambda=\Lambda\quad(8\pi G=1)\] one finds
\begin{equation}
\Lambda = (j-2q)H^2=(1-2q)H^2. \label{eq2.8}
\end{equation}
Here we have taken into account that $j=1$ in the SCM. Energy density of the cosmological constant does not depend on time, that is why explicit calculation of the time derivative $\dot\Lambda$ represents a natural test for correctness of the relation (\ref{eq2.8}). Using the expressions (\ref{hubble_paramters_time_derivatives}) for $\dot H$ and (\ref{alph.eq8}) for $\dot q$, it is easy to see that the condition $\dot\Lambda=0$ holds exactly:
\begin{align}
\nonumber \dot\Lambda=-2\dot q H^2+2(1-q)H\dot H=2H^3\left[ \left(1-2q^2-q\right) -(1-2q)(1+q)\right]=0.
\end{align}
As $\Lambda=const$, the right-hand side of the relation (\ref{eq2.8}) can be calculated at the parameters $q$ and $H$ values corresponding to any time moment. Using the current values of the parameters, for the cosmological constant $\Lambda$ and its relative density \[\Omega_{\Lambda0}\equiv\frac\Lambda{3H_0^2}\] at the present time one finds the following values:
\begin{align}
\label{eq2.10} &\Lambda=(1-2q_0)H_0^2,\\
\label{eq2.11} &\Omega_{\Lambda0}=\frac13(1-2q_0).
\end{align}
Of course, the result (\ref{eq2.11}) directly follows from the expression (\ref{eq1_111n1}) for the deceleration parameter in a flat Universe, however it was obtained here in an alternative way, which represents a universal method to obtain arbitrary model parameters.

Let us make sure that the above obtained expressions obey the flatness condition $k=0$, which can be reformulated in terms of the relative densities $\Omega_m$ and $\Omega_\Lambda$ in the form
\begin{equation}
\label{eq2.12} \Omega_m + \Omega_\Lambda=1.
\end{equation}
Using the above found solutions (\ref{eq2.11}), we obtain
\begin{equation}
\label{eq2.13} \Omega_m + \Omega_\Lambda=\frac23(1+q)+\frac13(1-2q)=1.
\end{equation}
Let us now perform a similar procedure for a Universe filled by the non-relativistic matter with the density \[\rho_m=\frac{M_m}{a^3}\] and radiation with density \[\rho_r=\frac{M_r}{a^4}\] which do not interact with each other.

We can represent the first Friedmann equation in the form
\begin{equation}
\label{eq2.14} \frac{\dot a^2}{a^2}+\frac k{a^2} = \frac{M_m}{a^3}+\frac{M_r}{a^4},\quad \frac{8\pi G}3=1.
\end{equation}
Taking time derivative twice, one finds
\begin{align}
\nonumber \ddot a =-\frac12\frac{M_m}{a^2}-\frac{M_r}{a^3},\\
\label{eq2.15} \dddot a =\frac{M_m}{a^3}\dot a+3\frac{M_r}{a^4}\dot a.
\end{align}
Using the definitions (\ref{eq2.3}) of the cosmographic parameters $q$ and $j$, one obtains
\begin{align}
\nonumber q&=\frac12A+B,\\
j&=A+3B,\quad A\equiv\frac{M_m}{a^3H^2},\quad B\equiv\frac{M_r}{a^4H^2}. \label{eq2.16}
\end{align}
One then finds
\begin{align}
\nonumber A&=-2j+6q,\\
\label{eq2.18} B&=j-2q.
\end{align}
The Friedmann equation (\ref{eq2.14}) in terms of the above introduced variables $A$ and $B$ takes on the form
\begin{equation}
\label{eq2.19} \frac k{a^2}=(A+B-1)H^2,
\end{equation}
or, in terms of the cosmographic parameters,
\begin{equation}
\label{eq2.20} k=a^2H^2(4q-j-1).
\end{equation}
Let us consider the limiting cases of the latter relation: the flat Universe solely filled with the non-relativistic matter and the flat Universe solely filled with radiation. In the former case, we have $A=1$, $B=0$, $k=0$, and $q=1/2$, therefore $j=1$ as it should be, because the considered Universe is nothing than a particular (matter-dominated) case of the SCM, where this relation holds exactly. In the latter case $A=0$, $B=1$, $q=1$, and $k=0$, then $j=3$.

The above considered procedure enables us to find the connections between the cosmographic parameters without any free parameters. Let us give an example for the above considered model of the Universe filled with the dark energy in form of the cosmological constant and the non-relativistic matter without interaction with each other.

Using the expression (\ref{eq2.2}) for $\dddot a$, we find
\begin{equation}
\label{eq2.21} \dddot a = \frac M3\frac{\ddot a}{a^3} -M\frac{\dot a^2}{a^4}+\frac\Lambda3\ddot a.
\end{equation}
For the parameter \[s\equiv a^3\frac{\ddddot a}{\dot a^4}\] one finds
\begin{equation}
\label{eq2.22} s=-(A+B)q-3A.
\end{equation}
Substitution of the quantities $A$ and $B$ into the latter equation gives
\begin{equation}
\label{eq2.23} s+2(q+j)+qj=0.
\end{equation}
The latter expression represents an ordinary differential equation of the fourth order, and it is absolutely equivalent to the first Friedmann equation. It was first obtained in the paper \cite{Dunajski:2008tg}.

The relation (\ref{eq2.23}) can be derived in an alternative way using the the fact that in the SCM \[\frac{dj}{dt}=0\] and therefore according (\ref{alph.eq8})
\begin{equation}
\label{eq2.24} \frac{dj}{dt}=H[s+j(2+3q)]=0.
\end{equation}
For the case $j=1$ the latter equation (\ref{eq2.24}) reproduces (\ref{eq2.23}).

Let us now perform the similar consideration for the case of the Chaplygin gas  with the equation of state
\[p=-\frac A\rho.\]
Solving the conservation equation, we find the energy density as a function of the scale factor
\begin{equation}
\label{eq2.25} \rho=\sqrt{A+Ba^{-6}},
\end{equation}
where $B$ is the integration constant. For small values of $a(t)$ density and pressure of the Chaplygin gas behave as the corresponding quantities of the non-relativistic matter:
\[\rho\propto a^{-3},\quad p=0,\]
and for large $a$ they mimic the cosmological constant:
\[\rho=const,\quad p=-\rho.\]
We are mostly interested in the intermediate region, where we can use the following approximation
\begin{equation}
\label{eq2.26} \rho=\sqrt A +\frac B{\sqrt{2A}}a^{-6}.
\end{equation}
We represent the first Friedmann equation in the form
\begin{equation}
\label{eq2.27} \dot a^2+k=\alpha a^2+\beta a^{-4},\quad \alpha\equiv\sqrt A,\quad \beta\equiv\frac B{\sqrt{2A}}.
\end{equation}
Consecutive differentiation of the latter equation allows us to evaluate the cosmographic parameters
\begin{align}
\nonumber q&=-\frac\alpha{H^2}+\frac{2\beta a^{-6}}{H^2}\equiv-M+N,\\
\label{eq2.28} j&=\frac\alpha{H^2}+\frac{10\beta a^{-6}}{H^2}\equiv M+5N,\\
\nonumber s&=-\frac{\alpha\ddot a}{aH^4}-\frac{60\beta a^{-6}}{H^2}+\frac{10\beta a^{-6}\ddot a}{aH^4}=-Mq-30N-5Nq.
\end{align}
Solving for $M$ and $N$, we find from the first two equations
\begin{align}
\nonumber M&=\frac16(j-5q),\\
\label{eq2.29} N&=\frac16(j+q).
\end{align}
Substituting $M$ and $N$ into the last equation of the system (\ref{eq2.28}), we finally get
\begin{equation}
\label{eq2.30} s+5(q+j)+qj=0.
\end{equation}
Of course, the latter relation is valid only in the limit \[\frac B{2\sqrt A}a^{-6}\ll1.\]
Similar procedure for the generalized Chaplygin gas with the equation of state \[p=-\frac A{\rho^\alpha}\] leads to the following:
\begin{equation}
s+(3\alpha+2)(q+j)+qj=0.
\end{equation}
We should note that for $\alpha=1$ we recover the above obtained result for the Chaplygin gas. If we want to exclude also the parameter $\alpha$ from the final equation, we have to perform additional differentiation of the Friedmann equation and introduce the additional cosmographic parameter \[l=\frac1a\frac{d^5a}{dt^5}\left(\frac1a\frac{da}{dt}\right)^{-5}.\] As the result we obtain
\begin{equation}
\label{eq2.32}
-2qs-2jq^2-lq-2sj-3sq^2-j^2q-lj+s^2-3q^2j-qsj+j^3-2j^2q^2=0.
\end{equation}
Deriving the latter relation we used the approximation analogous to the one used in (\ref{eq2.26}), and therefore it has a limited validity.

The relations between the time derivatives of density and pressure (\ref{alph.eq10}) obtained in Chapter 2 in some cases enable us to simplify considerably the procedure of determination of the model parameters without the use of the Friedmann equations. As an example, let us find the parameter $A$ for the Chaplygin gas with the state equation $p=-A/\rho$. Using the relations $p=-H^2(1-2q)$ and $\rho=3H^2$, we can obtain
\begin{equation}\label{eq2.33}
A=3H^4(1-2q)=3H_0^2(1-2q_0).
\end{equation}
The parameter A can be expressed also in terms of the higher cosmographic parameters. Using for example $\dot p=A\dot{\rho/\rho^2}$ and taking into account that $\dot{\rho=-6H^3(1+q)}$, we find
\begin{equation}\label{eq2.34}
A=\frac{\dot p}{\dot\rho}\rho^2=3H^4\frac{j-1}{q+1}.
\end{equation}
We obtained two different expressions for the same constant. It requires the existence of certain link between the cosmographic parameters that enter the two relations:
\begin{equation}\label{eq2.35}
1-2q=\frac{j-1}{1+q}\Rightarrow j=2-q-2q^2.
\end{equation}
Let us check that exactly this relation between the cosmographic parameters is realized in the Chaplygin gas model. Indeed, in the considered model
\begin{equation}
\frac{\dot p}{p}=-\frac{\dot{\rho}}{\rho}\Rightarrow1-2q=\frac{j-1}{1+q}\Rightarrow j=2-q-2q^2.
\end{equation}
We should stress that the relation $j=2-q-2q^2$ is not universal, it takes place only in the Chaplygin gas model.

The results obtained above can be generalized \cite{Dunajski:2008tg} on the case of $(n+1)$-dimensional uniform and isotropic Universe, filled with the cosmological constant $\Lambda$ and a component with the equation of state $p_m=w\rho_m$, which do not interact with each other. Let us express the curvature parameter $k$ in terms of the cosmographic parameters in this case.

The evolution equations for the considered Universe read
\begin{align}
\nonumber H^2&=\frac2{n(n-1)}\rho-\frac k{a^2},\quad 8\pi G=1,\\
\label{eq.2.37} \rho&=\rho_m+\Lambda,\\
\nonumber \dot\rho_m&+nH(\rho_m+p_m)=0.
\end{align}
Using the conservation equation, we find
\begin{equation}
\label{eq2.38}
\rho=\rho_0a^{-n(1+w)}+\Lambda.
\end{equation}
Substitution of this density into the first Friedmann equation gives
\begin{equation}
\label{eq2.39}
\ddot a=\frac2{n(n-1)}\rho_0a^{-n(1+w)+2}+\frac{2\Lambda}{n(n-1)}a^2-k.
\end{equation}
We introduce the notation
\begin{equation}
\label{eq2.40}
-n(1+w)=N,\quad \frac2{n(n-1)}=M.
\end{equation}
In this notation, the first Friedmann equation reads
\begin{equation}
\label{eq2.41}
\dot a^2=M\rho_0a^{N+2}+M\Lambda a^2-k.
\end{equation}
Differentiation of the latter equation with respect to time gives
\begin{align}
\nonumber \ddot a&=\frac12M\rho_0(N+2)a^{N+1}+M\Lambda a,\\
\label{eq2.42} \dddot a&=\frac{(N+1)(N+2)}2M\rho_0a^N\dot a+M\Lambda\dot a.
\end{align}
Using the cosmographic parameters, we can write
\begin{align}
\nonumber q&=-\frac{\ddot a a}{\dot a^2}=-\frac12M\rho_0(N+2)\frac{a^N}{H^2}-\frac{M\Lambda}{H^2},\\
\label{eq2.43} j&=\frac{\dddot a a^2}{dot a^3}=\frac{(N+1)(N+2)}2M\rho_0\frac{a^N}{H^2}+\frac{M\Lambda}{H^2}.
\end{align}
With the notations
\begin{equation}
A\equiv-M\rho_0(N+2)\frac{a^N}{H^2},\quad B\equiv\frac{M\Lambda}{H^2},
\end{equation}
the equation (\ref{eq2.43}) take on the form
\begin{align}
\nonumber q&=\frac12A-B,\\
\label{eq2.45} j&=-\frac{N+1}2A+B.
\end{align}
It then follows that
\begin{equation}\label{eq2.46}
A=-\frac2N (q+j),\quad B=-\frac1N[(N+1)q+j].
\end{equation}
In terms of the parameters $A$ and $B$ the first Friedmann equation takes on the form
\begin{equation}
\label{eq2.47} \frac k{a^2H^2}=-\frac1{N+2}A+B-1,
\end{equation}
or equivalently in terms of the cosmographic parameters $q$ and $j$
\begin{equation}
\label{eq2.48} k=a^2H^2\left\{\frac2{N(N+2)}(q+j)-\frac1N[(N+1)q+j]-1\right\}.
\end{equation}
For $n=3$, $w=0$ ($N=-3$) the latter expression for the spatial curvature reduces to
\begin{equation}
\label{eq2.49} k=a^2H^2(j-1)
\end{equation}
which reproduces the the above obtained equation (\ref{eq2.20}).

Let us now find the analogue of the equation (\ref{eq2.23}) for the multidimensional case. Using the above-obtained expression (\ref{eq2.42}) for $\dddot a$, we can find
\begin{equation}
\label{eq2.50} \ddddot a = \frac{N(N+1)(N+2)}2M\rho_0a^{N-1}\dot a^2+ \frac{(N+1)(N+2)}2M\rho_0a^N\ddot a +M\Lambda\dddot a.
\end{equation}
Taking into account the definition \[s=\frac{a^3}{\dot a^4},\] we obtain
\begin{equation}
\label{eq2.51} s=\left(\frac{N+1}2A-B\right)q-\frac{N(N+1)}2A.
\end{equation}
The parameters $A$ and $B$ are defined in the relations (\ref{eq2.46}). In terms of the cosmographic parameters we finally obtain
\begin{equation}
\label{eq2.52} s-\left\{ \frac2N(q+j)+\frac1N[(N+1)q+j]q-\frac6N(q+j)\right\}=0.
\end{equation}
The latter equation (\ref{eq2.52}) represents the first Friedmann equation expressed in terms of the cosmographic parameters for the $n$-dimensional case. For $N=-3$ ($n=3$, $w=0$) the above obtained result reproduces the equation (\ref{eq2.23}).

Let us now consider a more general setup for the cosmography of the two-component Universe (without the interaction between the components). Let the equations of state for the non-interacting components read
\begin{equation}
\label{eq2.53} p_A=w_A\rho_A,\quad p_B=w_B\rho_B.
\end{equation}
The energy densities $\rho_A$ and $\rho_B$ satisfy the following conservation equation
\begin{align}
\nonumber \dot\rho_A+3H(\rho_A+p_A)&=0,\\
\label{eq2.54} \dot\rho_B+3H(\rho_B+p_B)&=0.
\end{align}
Using the equations of state (\ref{eq2.53}) we find
\begin{align}
\nonumber \rho_A(a)&=\rho_0a^{-3(1+w_A)}\equiv\frac{A}{a^{\alpha+2}},\\
\label{eq2.55} \rho_B(a)&=\rho_0a^{-3(1+w_B)}\equiv\frac{B}{a^{\beta+2}},
\end{align}
where $A\equiv\rho_{0A}$, $B\equiv\rho_{0B}$, $\alpha\equiv3w_A+1$, $\beta\equiv3w_B+1$. Substitution of the latter solutions into the first Friedmann equation gives
\begin{equation}
\label{eq2.56} \dot a^2+k=\frac{A}{a^\alpha}+\frac{B}{a^\beta}.
\end{equation}
Taking twice the time derivative of the latter equation generates the following system of linear equations for the constants $A$ and $B$:
\begin{align}
\nonumber -2\frac{\ddot a}a&= 2H^2q=A\alpha a^{-\alpha-2}+B\beta a^{-\beta-2},\\
\label{eq2.57} 2\frac{\dddot a}a&=2H^2j=A\alpha(\alpha+1)a^{-\alpha-2}+B\beta(\beta+1)a^{-\beta-2}.
\end{align}
Solving the latter system we find
\begin{align}
\nonumber A&=\frac{2H^2[j-(\beta+1)q]}{\alpha(\alpha-\beta)}a^{\alpha+2},\\
\label{eq2.58} B&=\frac{2H^2[j-(\alpha+1)q]}{\beta(\beta-\alpha)}a^{\beta+2}.
\end{align}
Substitution of the above-found solutions into the Friedmann equation (\ref{eq2.56}) gives the following expression for the curvature parameter
\begin{equation}
\label{eq2.59} k=\frac{2a^2H^2}{\alpha\beta}[q(\alpha+\beta+1)-j]-1.
\end{equation}
It is easy to check that the relation (\ref{eq2.59}) correctly reproduces the above considered particular cases (\ref{eq2.7}) and (\ref{eq2.20}).
In order to establish the connections between the cosmographic parameters we need the additional differentiation of the Friedmann equation leading to
\begin{equation}
\label{eq2.60} \frac2H\left(\frac{\ddddot a}{\dot a}-\frac{a^{(V)}}{\dot a^2}\right) = 2H^2(s+qj) = -A\alpha(\alpha+1)(\alpha+2)a^{-\alpha-2} -B\beta(\beta+1)(\beta+2)a^{-\beta-2}.
\end{equation}
Using the definition of the cosmographic parameters (\ref{eq1_8}) and the above-found solution (\ref{eq2.58}), we can transform the equation (\ref{eq2.60}) to the following
\begin{equation}
\label{eq2.61} s+qj+(\alpha+\beta+3)j-q(\alpha+1)(\beta+1)=0.
\end{equation}
This differential equation represents a generalization of the equation (\ref{eq2.23}) on the case of the components with the equation of state $p_i=w_i\rho_i$, ($w_i=$ const). For the above considered case of the Universe filled with the dark energy in form of the cosmological constant and the non-relativistic matter, which do not interact with each other, we have $w_A=-1$, $w_B=0$, $\alpha=-2$, $\beta=1$, and we recover to the equation (\ref{eq2.23}).

The above considered procedure can be made more universal and efficient, if we start from the system of Friedman equations for the Hubble parameter $H$ and its time derivative $\dot H$. For the case of the multicomponent Universe this system of equations takes on the form
\begin{align}
\label{eq2.62} H^2=\frac13\rho-\frac k{a^2},&\quad\rho=\sum\limits_i\rho_i,\\
\label{eq2.63} \dot H=-\frac12(\rho+p)+\frac k{a^2},&\quad p=\sum\limits_i p_i.
\end{align}
Taking time derivatives of the equation (\ref{eq2.63}) as many times as needed, we obtain the system of equations including higher order time derivatives of the Hubble parameter $\ddot H$, $\dddot H$, and $\ddddot H$. This derivatives are immediately connected to the cosmographic parameters by the relations (\ref{hubble_paramters_time_derivatives}). Solving the corresponding system of equations we can find the free parameters of the model. As an example, let us apply this approach to the above considered case of the Universe filled with the dark energy in form of the cosmological constant and non-relativistic matter that do not interact with each other. The only difference from the SCM is that the spatial curvature is arbitrary. The model is described by the following system of Friedmann equations
\begin{align}
\label{eq2.64} H^2&=\frac13\rho_m+\frac13\Lambda-\frac k{a^2},\\
\label{eq2.65} \dot H&=-\frac12\rho_m+\frac k{a^2}.
\end{align}
Taking time derivative of the equation (\ref{eq2.65}), we find
\begin{equation}
\label{eq2.66} \frac k{a^2}=\frac{\ddot H}H+3\dot H.
\end{equation}
Substituting the expressions for $\dot H$ and $\ddot H$ that we already used many times above, we recover the result (\ref{eq2.7}). We should note that, unlike in the previous approach, we did not use here the solutions for the matter density as a function of the scale factor. Of course, for the case under consideration the explicit dependence can be trivially found, but a similar procedure can be applied to the models where it is really difficult to find the energy density as an explicit function of the scale factor.

Therefore, parameters of any model that satisfies the cosmological principle can be expressed in terms of the cosmographic parameters. Speaking more strictly, any one-parametric model can be described by two arbitrary cosmographic parameters. It is the most convenient to choose for them the current value of the Hubble parameter $H_0$ and the deceleration parameter. Thus, for example of the Chaplygin gas model with the equation of state $p=-A/\rho$ the parameter $A=3H^2_0(1-2q_0)$ (\ref{eq2.33}), and for the cosmological models with bulk viscosity $\zeta$ generating the additional negative pressure $p=-3H\zeta$ the parameter \[\zeta=\frac19H_0(1-2q_0)\] (see Sect.\ref{6.1}). The two-parametric models require already three cosmographic parameters: $H_0,\ q_0\to H_0,\ q_0,\ s_0$.

The proposed method enables us to significantly facilitate the testing of the cosmological models for the compatibility. It is crucial that the parameters of different models can be expressed through the unique parameter set. If some two models correspond to two non-overlapping regions of the cosmographic parameter space then the models are surely incompatible.

At last, as the proposed method allows us to fix the model parameters, we can answer the question: is it possible to achieve the goals the models was designed for? Thus, for example, a wide class of cosmological models was created to explain the accelerated expansion of the Universe without the use of the dark energy. All those models can realize the regime of the accelerated expansion only in certain regions of the parameter space. The proposed method enables us to clarify whether those regions correspond to the observed values of the cosmographic parameters.

The method is especially efficient in the cases where the dynamical variables of the model directly depend on the Hubble parameter (as for example for the case of the bulk friction with $p=-3H\zeta$ or in the models with the time-depending cosmological constant in the form $\Lambda(t)\to\Lambda(H)$). We show below some examples of the cosmological models (to be considered in details in the following Chapters), where the cosmographic analysis in terms of the Hubble parameter derivatives $\dot H,\ \ddot H,\ \dddot H,\ \ddddot H$ is definitely more efficient than the original approach of Dunajski and Gibbons \cite{Dunajski:2008tg}.

1. Dynamic models of vacuum energy, called "The running vacuum models" \cite{Sola:2015rra,Sola:2015csa,Lima:2007kk}, in which it is assumed that the density of vacuum energy depends on the Hubble parameter and its time derivatives,
	\begin{equation}
\label{eq2.67}\Lambda (H,\dot H) = C_0  + C_1 H^2  + C_2 \dot H + {\rm O}\left( {H^4 } \right)
\end{equation}
 	
2. Models with the creation of particles - "Creation model"\cite{Lima:2007kk,Pan:2013rha,Pan:2016jli,Chakraborty:2014fia}, in which negative pressure is generated in the process of creating particles due to the energy of the gravitational field. Models with the speed of creation of matter, depending on the Hubble parameter, allow us to describe the evolution of the Universe, starting with the post-inflation phase and ending with the current stage of accelerated expansion. In the most general model of this type,
	\begin{equation}
\label{eq2.68}\Gamma (H) = \Gamma _0  + \Gamma _1 H + \Gamma _2 H^2  + \Gamma _{ - 1} /H
\end{equation}

each member is responsible for a certain period of the evolution of the Universe.
3. Models that include bulk viscosity as a source of negative pressure \cite{Avelino:2008ph,Mostafapoor:2013jha}. As we will see below, the model parameters  $
\varsigma$
 can be expressed in terms of cosmographic parameters for both constant bulk viscosity and for more complex cases
	\begin{equation}
\label{eq2.69}\varsigma  = \varsigma _0  + H\varsigma _1  + \left( {\dot H + H^2 } \right)\varsigma _2  = \varsigma _0  + H\varsigma _1  + \left( {\dot H + H^2 } \right)\varsigma _2
\end{equation}
4. Cosmological models of a flat homogeneous and isotropic Universe, including additional entropic forces \cite{Komatsu:2014lsa}. It is assumed that the entropic forces depend on$H$
  and $\dot H$

\begin{align}
\nonumber H^2  = \frac{1}{3}\rho  + f(t),\quad f(t) = \alpha _1 H^2  + \alpha _2 \dot H, \\
\label{eq2.70} \dot H + H^2  =  - \frac{1}{6}\left( {\rho  + 3p} \right) + g(t),\quad g(t) = \beta _1 H^2  + \beta _2 \dot H
 \end{align}

Two types of models are considered. In the first, the entropy forces are included both in the first and in the second Friedmann equations ($\Lambda (H)$
-type models,$f(t) = g(t)$), and in the second case - only in the second Friedmann equation $f(t) = 0,$
 as is done in models with bulk viscosity.

\section{Reconstruction of cosmographic parameters}
The above-considered method to reconstruction of the cosmological model parameters based on usage of their kinematic characteristics assumes that we know current values of at least a few first cosmographic parameters. Reconstruction of the cosmographic parameters basing on a wide spectrum of the cosmological observational data represents an extremely complicated problem which includes a large number of processes of different nature taking place at a huge distance from the observer.

A trajectory in the configuration space represents the basic object of the kinematics. Kinematic characteristics of the trajectory (such as velocity, acceleration, etc.) help us to construct the dynamical model able to realize such trajectory. For example, Kepler orbits of the planets correspond to strictly defined dynamics (or at least sharply limit the number of the allowed models). In cosmology, the scale factor $a(t)$ serves as a trajectory $r(t)$. The scale factor cannot be immediately measured, but we can observe the cosmological phenomena characterized by different integrals of the scale factor. As we have seen above, the cosmographic parameters represent the coefficients of the Taylor series for the scale factor (see \ref{eq1_9}). The cosmographic parameters can in principle be determine by fitting of the integral characteristics with respect to the observations.

The most popular characteristic of that type is the distance (photometric, angular diameter or other) to the source of radiation. Application of decompositions of uh type faces an obvious problem: poor convergence of the series in the redshift $z$ for $z\le1$ and absence of the divergence for $z>1$. As was shown in \cite{Cattoen:2007sk}, this problem can be solved by introduction of the new variable
\begin{equation}
z\to y=\frac z{1+z}.
\end{equation}
When using the traditional redshift, the convergence radius is $R_z=1$. While the convergence range $[0,1]$ is formally the same for the $y$-variable, it corresponds to $[0,\infty]$ in terms of $z$. Therefore, the $y$-series remain within the convergence range even for the giant redshift values of the relic radiation
\begin{equation}
z\approx1100\to y=0.999.
\end{equation}
However, the necessity to take into account a large number of the series terms represents a problem even for the $y$-variable.

The procedure of determination of the cosmographic parameters can considered from another (perhaps more formal) point of view. The Hubble's law says nothing about the magnitude, sign or the very possibility of non-uniform expansion of the Universe. It is valid in the approximation which is insensitive to the acceleration. In order to investigate non-linear effects we need data for greater redshifts. If we could detect a deviation from the linearity in the Hubble diagram drawn on the basis of the observations, we would be able to determine the sign of the acceleration basing on the magnitude and sign of the deviation. If the deviation lies towards greater distances for a given redshift, then the acceleration is positive. The Hubble diagrams enable us to investigate even thiner geometrical features, in particular, to determine the deceleration parameter $q$, and even higher order cosmographic parameters $j$, $s$, $l$ (however with ever growing inaccuracy).

The distances are estimated by the luminosity of the source, under assumption that the considered population of sources represents the standard candles: an ensemble of objects with practically identical luminosity. Therefore the observed luminosity of such objects depends only on the distance to them. Supernova Ia bursts (exploding white dwarfs) give an example of the standard candles. As all the white dwarfs have almost identical mass, their luminosity is also practically the same. Their giant explosion power $10^{36}$W gives an additional bonus, as they can be observed up to very far distances.

The supernova bursts are seldom and occasional. In order to collect sufficient statistics we should control a significant part of the sky. The burst takes place during very limited time period, therefore it is crucial to detect the supernova as soon as possible in order to record evolution of its luminosity. Of course, the main question (whether we can consider the Ia supernovae as the standard candles) still remains open. At the beginning of the 90th in the USA two projects were launched in order to detect and analyze the SNE Ia bursts: they were called the SuperNova Cosmology Project and High-Z SuperNova Search. Both of them analyzed the so called golden set of the SNe Ia, containing 157 well studied SNe Ia with the redshift in the range $0.1<z<1.76$. In 1998-1999 the results of the two groups \cite{Riess:1998cb,Perlmutter:1998np} allowed to establish the fact of the accelerated expansion of the Universe, that totally changed the modern cosmology, and all the physics in general. The results were multiply repeated during the next decade with ever growing statistics. The main conclusion always remained the same: the Universe made transition from the decelerated expansion to the accelerated one comparably recently (at $z\approx0.5$).

The method based on the observation of the supernovae bursts remains the leader even today. But it occurred to have the competitors. A promising and absolutely independent substitute (rather than a surrogate) is the observation of the angular diameter distances $D_A(z)$ for a given set of distant objects. A combination of the Syunyaev-Zeldovich effect \cite{Zeldovich:1969ff} and measurements of the surface brightness in the X-ray range enables us to measure the angular diameter distance for the galactic clusters \cite{Bonamente:2005ct}. The Syunyaev-Zeldovich effect represents a small distortion of the CMB spectrum due to the inverse Compton scattering of the CMB photons that go through a population of the hot electrons.  Observations of the temperature fluctuations in the relic radiation spectrum of the galactic clusters, and the observations in the X-ray range give us the possibility to obtain the dependence $D_A(z)$ in an independent way. Therefore, $D_A(z)$ enables us to reconstruct the dynamics of the Universe independently from the $d_L(z)$.

Let us now consider in details this new promising possibility to research the history of the cosmological expansion of the Universe. Recall that for the standard candles the cosmologists initially used the Cepheids: the stars with the intensity proportional to the period of the brightness variation. A classical example of a Cepheid is the Polar star, which is the brightest and nearest to Earth, with the variable luminosity period equal to $3.97$ days. Cepheids are good standard candles for galactic distances. They enabled us to determine dimensions of our Galaxy and the distance to our nearest neighbor---Andromeda galaxy. In order to research dynamics of the Universe, we have to consider much larger scales and thus much more powerful standard candles. Remember, that the cosmological principle (which all the above-considered equations of the Universe dynamics are based on) postulating homogeneity and isotropy of the Universe, works on the scales large than 100 Mpc. Using Ia supernovae as the standard candles of much higher power enabled us to penetrate much deeper into the history of the Universe. However, those standard candles also had very limited potential. Up to the present, time we could observe Ia supernovae only for $z<2$, however, reliable reconstruction of the cosmological expansion history requires to consider much higher redshift values, and thus much more powerful standard candles. It turned out that such objects are available: it is the so-called gamma-ray-bursts (GRB) \cite{vedrenne2009gamma,Kumar:2014upa}: giant energy outbursts of the explosive type with duration from three to a hundred seconds observed in the hardest part of the electromagnetic spectrum. The energy amount $10^{54}$ erg radiated in the GRB is by an order of magnitude higher than that of a supernova. It is comparable to the Sun rest mass! The events generating the GRB are so powerful that can be sometimes observed by naked eye, though they occur at the distances of the order of billions light years from the Earth. The energy yield takes place in form of a collimated flow or a jet. That is because of the jets that we can see only a tiny fraction of all the gamma-bursts occurring on the Universe. Distribution of the GRB over the burst duration is clearly bi-modal.

The short GRB are possibly caused with a merger of neutron stars, or a neutron star and a black hole. The longer events are presumably connected with a formation of a black hole in a collapse of a massive star (more than 25 Solar mass) with significant angular momentum---it is the so-called collapsar model. We can prove the possibility to use the GRB as a standard candle basing on the so-called 'Amati relation' \cite{2005NCimCamati} which relates the peak frequency of the burst to its total energy---this is an exact analogue of the period-luminosity relation for the Cepheids. GRB as the standard candles yet have limited application because of the great dispersion of their characteristics, however, this method is potentially very attractive as it gives us a possibility to advance in considerably higher redshift range.

We can obtain information on the cosmographic parameters using the so-called baryon acoustic oscillations \cite{Bernardeau:2001qr,Crocce:2007dt}. Similar to the standard candles, they can be used as the standard rulers, i.e. the objects with fixed linear dimensions. The baryon acoustic oscillations are regular periodic fluctuations in density of the observed baryon matter. These fluctuations generate the potential wells which are responsible for a certain type of CMB anisotropy that can be observed. As we know the size of the sound horizon at the recombination moment, we can determine the current distance to the last scattering surface. As we have seen above, any distance in the expanding Universe can be represented in form of series in the redshift, and the coefficients of the series are the cosmographic parameters.

At last, one more elegant method to determine the cosmographic parameters relies on the decomposition (\ref{alph.eq2})
\begin{equation}\label{eq2.74}
H(z)=H_0\left(1+(1+q_0)z+\frac12(j_0-q_0^2)z^2+\frac16(3q_0^2+3q_0^3-4q_0j_0-3j_0-s)z^3\right)
\end{equation}
If we could reconstruct the series (\ref{eq2.74}) from the observational data, then the problem of determination of the cosmographic parameters would reduce to solving a system of non-linear algebraic equations. Such possibilities really exist \cite{Farooq:2013hq}. In particular, it was recently shown that the so-called shining red galaxies give us an additional possibility of direct measurement of the expansion rate \cite{Jimenez:2001gg,Zhang:2010ic,Ma:2010mr}. The idea is to reconstruct the Hubble parameter from the time derivative of the redshift:
\begin{equation}
H(z)=-\frac1{1+z}\frac{dz}{dt}.
\end{equation}
The derivative can be found by measuring the 'age difference' between two passively evolving galaxies at different yet close redshifts. The method was already realized for $0.1<z<1.75$. We should stress that this range includes the transition region under interest which has $z\approx0.5$. The results of the analysis of the available data \cite{Zhang:2010ic} agree with the data obtained from the supernovae and the angular diameter distance. In the nearest future it is expected to obtain an array of 2000 passively evolving galaxies in the range $0<z<1.5$. These observations would enable us to obtain 1000 $H(z)$ values with $15\%$ accuracy, provided the galaxy ages are known with $10\%$ accuracy.

The Tables \ref{t1}-\ref{t3} show the values of the cosmographic parameters $h\equiv H_0/100 \mathrm{\ km\ sec}^{-1} \mathrm{Mpc}^{-1}$, $q_0$, $j_0$, $s_0$, obtained \cite{Demianski:2016dsa} by the combination of the four above-described observational methods.
\begin{table}[ph]
\tbl{Cosmographic parameters reconstructed from the data on Ia SN, BAO and $H(z)$ \cite{Demianski:2016dsa}.}
{
\begin{tabular}{c|c|c|c|c}
\hline\hline
Parameter & $h$ & $q_0$ & $j_0$ & $s_0$ \\
\hline\hline
Best fit & 0.74 & -0.48 & 0.68 & -0.51 \\
\hline
Mean & 0.74 & -0.48 & 0.65 & -6.8 \\
\hline
$2\sigma$ & (0.68, 0.72) & (-0.5, -0.38) & (0.29, 0.98) & (-1.33, -0.53) \\
\hline
\end{tabular}}
\label{t1}
\end{table}

\begin{table}
\tbl{Cosmographic parameters reconstructed from the data on GRB, BAO and $H(z)$ \cite{Demianski:2016dsa}.}
{\begin{tabular}{c|c|c|c|c}
\hline\hline
Parameter & $h$ & $q_0$ & $j_0$ & $s_0$ \\
\hline\hline
Best fit & 0.67 & -0.14 & 0.6 & -5.55 \\
\hline
Mean & 0.67 & -0.14 & 0.6 & -5.55 \\
\hline
$2\sigma$ & (0.66, 0.73) & (-0.15, -0.14) & (0.58, 0.62) & (-5.7, 6.1) \\
\hline
\end{tabular}\label{t2}}
\end{table}

\begin{table}
\tbl{Cosmographic parameters reconstructed from the data on Ia SN, GRB and BAO \cite{Demianski:2016dsa}.}
{\begin{tabular}{c|c|c|c|c}
\hline\hline
Parameter & $h$ & $q_0$ & $j_0$ & $s_0$ \\
\hline\hline
Best fit & 0.72 & -0.6 & 0.7 & -0.36 \\
\hline
Mean & 0.72 & -0.6 & 0.7 & -0.37 \\
\hline
$2\sigma$ & (0.67, 0.73) & (-0.62, -0.55) & (0.69, 0.73) & (-0.4, 5) \\
\hline
\end{tabular}\label{t3}}
\end{table}

Even a glance on the Tables \ref{t1}-\ref{t3}, enables us to make the following conclusions. The values of the cosmographic parameters in the Tables \ref{t1}-\ref{t3} rather well agree to each other, but the deceleration parameter value (which is a crucial cosmological parameter) in the Table \ref{t2} (-0.14) remarkably both from the values (-0.6) in the other two Tables and from the value ($\approx$0.55) obtained in the SCM. Similarly for the parameter $s_0$: in the Tables \ref{t1},\ref{t2} this parameter is close to the value expected in the SCM:\[s_0=1-\frac92\Omega_{m0}\approx-0.325.\] The Table \ref{t2} is the only one that is calculated disregarding the supernova data. This is one more time that their crucial role as very reliable standard candles should be stressed.

We should make an important remark: the traditional kinematics of the classical mechanics is model-free---you need only a clock and a ruler to determine velocities or accelerations. Unfortunately, when determining the cosmographic parameters, we lack both the appropriate ruler and the clock to measure the cosmological time. We need a model to calculate both the distances and the time intervals in the expanding Universe. But all the above-obtained formulae are exact and valid for arbitrary values of the cosmograpic parameters. They are waiting for the Cosmic Concordance with respect to the cosmographic parameter values.

To avoid the confusion, we should stress that there are two sets of the cosmographic parameters. The first set $H,q,j\dots\equiv(CP)$ is generated by the decomposition of the real scale factor $a(t)$, and the coefficients in the decomposition can be found analyzing the observations as was shown above. Elements of the set $(CP)_0$ obey a number of universal (model-free) relations (see Sect.\ref{alphabet}), which contain time derivatives (or derivatives with respect to the redshift). For example,
\begin{align}
\nonumber j&=-q+2q(1+q)+(1+z)\frac{dq}{dz},\\
\label{eq2.76} s&=j-3j(1+q)-(1+z)\frac{dj}{dz}.
\end{align}
Each particular cosmological model that has a solution $a^*(t)$ generates an alternative paramter set $H^*,q^*,j^*\dots\equiv(CP)^*$. The proximity (which of course requires a detailed definition) of the sets $(CP)$ and $(CP)^*$ tells us that the corresponding model is adequate.

As we have already mentioned, all the cosmographic parameters $(CP)^*$ can be constructed under the assumption that a few first cosmographic parameters are known. The proposed method to determine the model cosmographic parameters is simple and universal. We have shown above that the first Friedmann equation (after exclusion of the model parameters) can be represented solely in terms of the cosmographic parameters. The equations (\ref{eq2.23}), (\ref{eq2.32}), and (\ref{eq2.51}) give an example of the relations of such type. Multiple differentiation of those equations with respect to time followed by expression of the time derivatives of the cosmographic parameters making use of the relations (\ref{alph.eq8}) directly in terms of the lower-order cosmographic parameters enables us to express any cosmographic parameter through the several given ones.

Let illustrate the latter notion by the following example of the SCM. Let us assume that the two first cosmographic parameters are given:
\begin{equation}\label{eq2.77}
q=-1+\frac32\Omega_m,\ j=1.
\end{equation}
From (\ref{eq2.23}) we find
\begin{equation}\label{eq2.78}
s=-3q-2=1-\frac92\Omega_m.
\end{equation}
Taking derivative of the equation (\ref{eq2.23}) with respect to time and making use of the relation for the time derivatives of the cosmographic parameters (\ref{alph.eq8}), we obtain
\begin{equation}\label{eq2.79}
\dot s=-2\dot j-\dot q\Rightarrow l=1+3\Omega_m+\frac92\Omega_m^2.
\end{equation}
A similar procedure can be used in any other model which satisfies the Cosmological Principle. 
\section{Cosmography of one-component models without the dark energy}
According to the current cosmological paradigm, we live in a flat accelerated expanding Universe. However, the physical origin of the cosmic acceleration is still the greatest mystery. The Universe filled with "ordinary" components (matter and radiation) should eventually slow down the expansion. Any attempt to explain the observed accelerated expansion of the Universe, remaining within the framework of General Relativity, can be done only by modifying either the left or the right side of the Einstein equations. SCM achieves agreement with cosmological observations by including in the energy-momentum tensor a component of an unknown nature: dark energy and dark matter. At the phenomenological level, the introduction of dark components can be avoided by modifying the Friedmann equations, that is, changing the direction of the evolution of the ordinary components (matter and radiation) in the required direction. It turned out that the desired goal can be achieved in various ways, including: the replacement of the ideal fluid by a viscous one, the introduction of sources into the conservation equation, the transformation of the equations of state, and many others. This Implementation of models of such type requires a lot of additional parameters. In this section, we express the parameters of a number of popular models through a universal set of cosmological parameters.
\subsection{\label{6.1}Kinematics of cosmological models with bulk viscosity}
As far back as the 70s of the last century it became clear that one of the sources of negative pressure necessary for the accelerated expansion of the Universe is the bulk viscosity \cite{Murphy:1973zz, Padmanabhan:1987dg}. The bulk viscosity (sometimes called the "second viscosity") is a consequence of a violation of local thermodynamic equilibrium. In the context of the dynamics of the Universe, the bulk viscosity can arise because the thermodynamic equilibrium simply does not have time to recover when the expansion is sufficiently rapid. The additional pressure that arises when the thermodynamic equilibrium is disturbed tends to return the system to the equilibrium. The pressure generated by bulk viscosity decreases as we approach the thermodynamic equilibrium. In the last decade, bulk viscosity as an alternative to dark energy was considered in the papers \cite{Fabris:2005ts,Li:2009mf,HipolitoRicaldi:2010mf,Avelino:2008ph,Avelino:2010pb,Sasidharan:2014wqa,Sasidharan:2015ihq,Mohan:2017poq,Mostafapoor:2013jha}.

An alternative approach to introduce the bulk viscosity is to treat the dark matter and the dark energy as a single substance with the equation of state $p=p(\rho)$ or $p(H)$. We note that for a planar Universe these equations of state are equivalent. Models of such a unifying type include the Chaplygin gas model mentioned above with the equation of state
\begin{equation}\label{eq3_1}
p(\rho)=-\frac A{\rho^\alpha}.
\end{equation}
where the constant $A>0$ and $0<\alpha<1$.

At the phenomenological level, the additional pressure generated by bulk viscosity can be included in the energy-momentum tensor of an ideal fluid, transforming it into a non-ideal \cite{Zimdahl:1996fj}
\begin{equation}\label{eq3_2}
T_{\mu\nu}=(\rho+p-3H\xi)u_\mu u_\nu+(p-3H\xi)g_{\mu\nu}
\end{equation}
where $\xi$ is the bulk viscosity parameter, which will be referred to as bulk viscosity for simplicity. This parameter enters into the Navier-Stokes equation, which describes the motion of a viscous fluid with non-relativistic velocities. The role of bulk viscosity reduces to generation of the additional pressure $-3H\xi$.

The bulk viscosity, generally speaking, depends on the rate of cosmological expansion. An approximate form of this dependence was found in \cite{Ren:2005nw}, based on the following considerations. The bulk viscosity is used for the phenomenological description of the unknown properties of the dark components. Since the equation of state of the dynamic dark energy depends on the rate of expansion of the Universe, then the bulk viscosity simulating the properties of the dark components should change as the Universe evolves. Capozziello et al. \cite{Capozziello:2005pa} showed that the equation of state with constant coefficients
\begin{equation}\label{eq3_3}
p=p_0+w\rho+w_HH+w_{H^2}H^2+w_{dH}\dot H
\end{equation}
describes quite well the complete history of the cosmological expansion. Ren and Ming \cite{Ren:2005nw} found that the bulk viscosity
\begin{equation}\label{eq3_4}
\xi=\xi_0+\xi_1H+\xi_2\frac{\ddot a}a
\end{equation}
is equivalent to the equation of state (\ref{eq3_3}). The purpose of this section, using the approach proposed in the previous chapter, is to express the constants $\xi_0$, $\xi_1$, $\xi_2$ through the cosmological parameters.

Let us start with the model of a flat Universe filled with a single component - the non-relativistic matter (both baryon and dark) with the energy density $\rho_m$ and constant bulk viscosity ($\xi=\mathrm{const}$) \cite{Avelino:2008ph}. The conservation equation and the first Friedmann equation in this case have the form
\begin{align}
\nonumber
& a\frac{d\rho_m}{da}-3(3H\xi-\rho_m)=0,\\
\label{eq3_5}
& H^2=\frac{8\pi G}3\rho_m.
\end{align}
Excluding the Hubble parameter and changing the variables from the scale factor to the redshift, we find
\begin{equation}\label{eq3_6}
(1+z)\frac{d\rho_m}{dz}-3\rho_m+\gamma\rho_m^{1/2}=0, \quad \gamma\equiv9\sqrt{\frac{8\pi G}{3\xi}}
\end{equation}
The solution of this equation is
\begin{equation}\label{eq3_7}
\rho_m(z)=\left[\frac\gamma3+\left(\rho_{m0}^{1/2}-\frac\gamma3\right)(1+z)^{3/2}\right]^2.
\end{equation}
Substituting the above found solution $\rho_m(z)$ into the first Friedmann equation, we obtain for the Hubble parameter
\begin{equation}\label{eq3_8}
H(z) = H_0\left[\frac{\bar{\xi}}{3}+\left(\Omega_{m0}^{1/2}-\frac{\bar{\xi}}{3}\right)(1+z)^{3/2}\right],\quad\bar{\xi}\equiv\frac{24\pi G}{H_0}\xi.
\end{equation}
For the one-component flat Universe $\Omega_{m0}=1$ and, consequently,
\begin{equation}\label{eq3_9}
H(z) =\frac13H_0\left[\bar{\xi}+(3-\bar{\xi})(1+z)^{3/2}\right].
\end{equation}
The obtained expression allows us to find the scale factor as a function of time. To do this, we transform the expression for the Hubble parameter
\begin{equation}\label{eq3_10}
H(a) =\frac13H_0\left(\frac{\bar{\xi}a^{3/2}+3-\bar{\xi}}{a^{3/2}}\right).
\end{equation}
to
\begin{equation}\label{eq3_11}
H_0(t-t_0)=3\int\limits_{1}^{a}\frac{a'^{1/2}}{\bar{\xi}a'^{3/2}+3-\bar{\xi}}da'.
\end{equation}
For $\xi\ne0$ and $\bar{\xi}a^{3/2}+3-\bar{\xi}>0$ ($\bar{\xi}a^{3/2}+3-\bar{\xi}<0$ implies $H<0$ and contradicts the observations) we find
\begin{equation}\label{eq3_12}
a(t)=\left[\frac{3\exp\left(\frac12\bar{\xi}H_0(t-t_0)\right)-3+\bar{\xi}}{\bar{\xi}}\right]^{2/3}.
\end{equation}
We see that in the asymptotics $t\to\infty$ in the interval $0<\bar{\xi}<3$ this solution demonstrates the de Sitter behavior \[a(t)\propto\exp\left(\frac{\bar{\xi}}{3}H_0(t-t_0)\right)\] On the other hand, for all values of the bulk viscosity in this interval in the considered model ($\xi=\mathrm{const}$), the Universe has experienced a "Big Bang" in the past. Let us determine how far this event took place in the past. The time of the Big Bang $t_{BB}$, determined by the condition $a(t_{BB})=0$, is
\begin{equation}\label{eq3_13}
t_{BB}=t_0+\frac2{\bar{\xi}H_0}\ln\left(1-\frac{\bar{\xi}}3\right)
\end{equation}
This expression for zero bulk viscosity ($\bar{\xi}=0$) correctly reproduces the lifetime of the Universe filled with non-relativistic matter
\begin{equation}\label{eq3_14}
H_0(t_0-t_{BB})=\frac23.
\end{equation}
Let us now proceed to calculate the deceleration parameter $q(a,\xi)$. The second Friedmann equation for the considered model has the form
\begin{equation}\label{eq3_15}
\frac{\ddot a}a=-\frac{4\pi G}3(\rho_m-9\xi H).
\end{equation}
Transforming variables to \[\bar{\xi}\equiv\frac{24\pi G}{H_0}\xi\] and substituting \[\rho_m=\frac3{8\pi G}\], we obtain
\begin{equation}\label{eq3_16}
\frac{\ddot a}a=-\frac12(\bar{\xi}H_0-H)H.
\end{equation}
Using the definition \[q(a)=-\frac{\ddot a}{aH^2}\] and expression \ref{eq3_10} for the Hubble parameter, we find
\begin{equation}\label{eq3_17}
q(a,\bar{\xi})=\frac12\left[\frac{3-\bar{\xi}(1+2a^{3/2})}{3-\bar{\xi}(1-a^{3/2})}\right].
\end{equation}
The expression obtained above allows us to classify the kinematics of the expansion of the Universe depending on the value of the bulk viscosity parameter $\bar{\xi}(\xi)$.
\begin{enumerate}
\item If $\bar{\xi}=0$ then $q=1/2$, that corresponds to the Universe with domination of matter, which has zero bulk viscosity.
\item If $\bar{\xi}=3$ then $q=-1$, that corresponds to the de Sitter's regime.
\item In the interval $0<\bar{\xi}<3$ the deceleration parameter is a monotonically decreasing function from the value $q(0)=1/2$ to $q(\infty)=-1$. The value of the scale factor at which the transition from decelerated expansion to accelerated one ($q(a_t)=0$) takes place is equal to
\begin{equation}\label{eq3_18}
a_t=\left[\frac{3-\bar{\xi}}{2\bar{\xi}}\right]^{2/3}.
\end{equation}
It is interesting to note that for $\xi=1$ the transition from decelerated expansion to accelerated one occurs today: $a_t=1$.
\item If $\bar{\xi}>3$ the expansion of the Universe is always accelerated, there are no periods of decelerated expansion.
\end{enumerate}
Using \ref{eq3_17}, we find the current value of the deceleration parameter
\begin{equation}\label{eq3_19}
q(a=1,\bar{\xi})=\frac{1-\bar{\xi}}{2}.
\end{equation}
The latter relationship can be used to solve the problem posed: to express the bulk viscosity parameter through the cosmographic parameters
\begin{equation}\label{eq3_20}
\bar{\xi}=2q_0-1\to\xi=\frac13H_0(1-2q_0).
\end{equation}
However, this solution is valid only for the simplest case $\xi=\mathrm{const}$. Only in this case there exists an analytic solution (\ref{eq3_12}) for the scale factor, with the help of which all subsequent relations were obtained. Below we show that the approach proposed in the previous chapter allows one to express bulk viscosity through cosmological parameters for a more general case of variable bulk viscosity (\ref{eq3_4}), obtaining precise analytical formulas and spending significantly less computational effort.

To see this, we consider the system of Friedmann equations in the form ($8\pi G=1$)
\begin{align}
\nonumber
H^2&=\frac13\rho,\\
\label{eq3_21}
\dot H &=-\frac12(\rho-3H\xi).
\end{align}
The second equation is obtained using the conservation equation
\begin{equation}\label{eq3_22}
\bar{\xi}=2q_0-1\to\xi=\frac13H_0(1-2q_0).
\end{equation}
The system \ref{eq3_21} allows us to obtain for the bulk viscosity parameter the expression
\begin{equation}\label{eq3_23}
\xi=\frac13H(1-2q).
\end{equation}
If $\xi=\mathrm{const}$, then the right-hand side of (\ref{eq3_23}) should not depend on time. Let us verify this in the following way.

We consider the system of equations generated by successive differentiation of the first Friedmann equation
\begin{align}
\label{eq3_24a}
H^2&=\frac13\rho\\
\label{eq3_24b}
\frac23\dot H&=-H^2+\dot H\xi\\
\label{eq3_24c}
\frac23\ddot H&=-2H\dot H+\dot H\xi.
\end{align}
As we saw above, the second equation allows us to find the bulk viscosity parameter $\xi$. Substituting the expression (\ref{eq3_23}) into the third equation, we obtain an equation containing only cosmographic parameters
\begin{equation}\label{eq3_25}
j-\frac12(1+q)-q^2=0.
\end{equation}
This equation represents a third-order ODE ($j=\dddot a/(aH^3)$) for a scale factor as a function of time. However, we are interested not so much in solutions of this equation, as in the relationship between the cosmographic parameters that can be obtained without resorting to these solutions. In particular, the relation (\ref{eq3_25}) allows us to express (within the frames of the model under consideration) any cosmographic parameter through the deceleration parameter. The relationship between the parameters $j$ and $q$ is directly given by this relation.

Using the data given in the previous section for the current value of the deceleration parameter $q_0\approx-0.48$ and $q_0\approx-0.6$ (see Table \ref{t1}-\ref{t3}), we find for the parameter $j$ the values of $j_0\approx0.49$ and $j_0\approx0.56$. These values agree with the observations $j_0\approx0.68$ and $j_0\approx0.7$, however, they are obtained with the aid of simple exact analytics.

The links of the deceleration parameter with higher cosmographic parameters can be obtained by successive differentiation with respect to time of the relation (\ref{eq3_25}) and subsequent substitution of time derivatives of the cosmological parameters (\ref{alph.eq8}). Let us calculate as an example the cosmographic parameter $s$,
\begin{align}
\nonumber \dot j-\frac12\dot q-2q\dot q=0,\\
\nonumber \ddot q=-H(j-2q^2-q),\\
\nonumber \ddot j=H[s+j(2+3q)],\\
\label{eq3_26}
s=\frac52j-qj-5j^2+\frac12q.
\end{align}
Substitution of $j=\frac12(1+q)+q^2$ into the latter expression allows us to achieve the desired result. We emphasize that the relations (\ref{eq3_25}) and (\ref{eq3_26}) are obtained without using solutions for the scale factor and they are exact.

Returning to the relation (\ref{eq3_23}), it is easy to see that condition
\begin{equation}\label{eq3_27}
\frac{d\xi}{dt}=0=\frac{d}{dt}H(1-2q)
\end{equation}
reduces to (\ref{eq3_25}), which allows us to consider $H(1-2q)$ as one of the integrals of the ODE (\ref{eq3_25}). This allows us to calculate the right-hand side of (\ref{eq3_23}) at any time. For the current time, we get the result exactly the same as (\ref{eq3_20}).

Let's make an important remark. Successively differentiating the Friedmann equation, we can use any pair of equations to find the bulk viscosity. The higher the derivatives of the Hubble parameter will be used, the higher the cosmological parameters will enter into the formula for bulk viscosity. Of course, the solutions obtained by using any pair of equations must coincide identically. Such an equivalence can be realized only if there is a connection between the cosmographic parameters.

Let us demonstrate the correctness of this assertion on the example of the model under consideration with a constant bulk viscosity. Expression (\ref{eq3_23}) for the bulk viscosity parameter was obtained using the equation including the first order time derivative of the Hubble parameter $\dot H$ (\ref{eq3_24b}). However, an expression for this parameter can be obtained using the equation for an arbitrary time derivative $H^{(n)}$. In particular, from equation (\ref{eq3_24c}) we find
\[
\xi=\frac23H\frac{j-1}{q+1}.
\]
Equating the expressions (\ref{eq3_28}) and (\ref{eq3_23}), we find that for the identical coincidence of these expressions, the connection (\ref{eq3_25}) between the deceleration parameter $q$ and the parameter $j$ is necessary. In order to verify the correctness of the relation obtained, we recall that $j=1$ in SCM. The substitution $j=1$ into (\ref{eq3_28}) leads to $\xi=0$, which is natural, since there is no viscosity in the SCM.

Introduction of the viscosity means a deviation from SCM and as a consequence in this case $j\ne1$. Using the observational data, it was shown in \cite{Kowalski:2008ez} that the value \[\xi=\frac19H_0\bar{\xi},\] where
\begin{equation}\label{eq3_28}
\bar\xi=\frac{2\pi G}{H_0}\xi=1.922\pm0.089
\end{equation}
represents the best estimate for the bulk viscosity. Such a value of bulk viscosity requires the fulfillment of the relationship $j_0+q_0\approx0$. This result agrees well with the latest estimates \cite{Luongo:2015zaa}, according to which
\begin{equation}\label{recent_estimates}
q_0=-0.527_{-0.088}^{+0.093},\quad j_0=0.501_{-0.527}^{+0.558}
\end{equation}
A similar relationship between these parameters occurs for the values of these parameters obtained above in the cosmographic approach.

Cosmographic parameters reconstructed by GRB, BAO and $H(z)$ are
\[q_0\approx-0.48,\quad j_0\approx0.49.\]
Cosmographic parameters reconstructed by SN 1a, GRB and BAO data are
\[q_0\approx-0.6,\quad j_0\approx0.56.\]
Above we considered the simplest version of bulk viscosity, when $\xi=\mathrm{const}$. The proposed method proves to be effective also for a more complex case, when the bulk viscosity varies with the expansion of the Universe \cite{Mostafapoor:2013jha}
\begin{equation}\label{eq3_29}
\xi=\xi_0+\xi_1H.
\end{equation}
In this case, the system of equations for determination of the parameters $\xi_0$ and $\xi_1$ has the form (in the following formulas we use the condition $8\pi G/3=1$)
\begin{align}
\nonumber
\frac{\dot H}{H^2} &=-\frac32+\frac32\frac{\xi_0}{H}+\frac32\xi_1,\\
\label{eq3_30}
\frac{\ddot H}{H\dot H} &=-3+\frac32\frac{\xi_0}{H}+3\xi_1.\\
\end{align}
Taking into account that \[\frac{\dot H}{H^2}=-(1+q)\] and \[\frac{\ddot H}{h\dot H}=-\frac{j+3q+2}{1+q},\] we obtain
\begin{align}
\nonumber
\xi_0 &=\frac23H\frac{j-q(2q+1)}{q+1},\\
\label{eq3_31}
\xi_1 &=\frac13\frac{-2j+2q^2+q+1}{q+1}.\\
\end{align}
Let's check the obtained result. The condition $\xi_1=0$ is transformed into the condition $2j=2q^2+q+1$. Substituting this value for $j$ into the expression for $\xi_0$, we will see that
\begin{equation}\label{eq3_32}
\xi_0=\frac29H\frac{j-q(2q+1)}{q+1}\to\frac19H(1-2q).
\end{equation}
This precisely reproduces the result \ref{eq3_23} obtained above for the case $\xi=\mathrm{const}$ (taking into account the transition from the condition $8\pi G=1$ to the condition $8\pi G/3=1$).
Using a similar procedure for the model
\begin{equation}\label{eq3_33}
\xi=\xi_0+\xi_1H+\frac{\ddot a}{a}\xi_2=\xi_0+\xi_1H+(\dot H+H^2)\xi_2
\end{equation}
we find
\begin{align}
\nonumber
\frac{\xi_0}{H}=-\frac29q-\frac49\frac{(1+q)(j-q^2)^2}{j^2+2jq-q^3+s+qs},\\
\nonumber
\xi_1=-\frac19\left(1+2q-\frac{4(1+q)(j+q)(q^2-j)}{j^2+2jq-q^3+s+qs}\right),\\
\label{eq3_34}
H\xi_2=\frac19\frac{j^2+2j(1+q(3+q))+s+q(s-q(2+q)(1+2q)}{j^2+2jq-q^3+s+qs}.
\end{align}
Let us dwell in more detail on the physical sense of the dependence of the volume viscosity parameter on the rate of expansion of the Universe, given by the Hubble parameter. The bulk viscosity leads to the generation of additional pressure
\begin{equation}\label{eq3_35}
P_{eff}=p+\Pi.
\end{equation}
where $p$ is the "ordinary" liquid pressure given by the equation of state $p=w\rho$, and $\Pi$ is the pressure due to the bulk viscosity. According to Israel-Stewart theory \cite{Israel:1976tn,Israel:1976PhLA}, it satisfies equation
\begin{equation}\label{eq3_36}
\tau\dot\Pi+\Pi=-3H\xi-\frac12\tau\Pi\left(3H+\frac{\dot{\tau}}{\tau}-\frac{\dot{\xi}}{\xi}-\frac{\dot T}{T}\right)
\end{equation}
Here $\tau$, $\xi$ and $T$ stand for the relaxation time, bulk viscosity and temperature, respectively. It is assumed that all these quantities are functions of the density of the liquid \cite{Maartens:1995wt}
\begin{align}
\nonumber
\tau&=\alpha\rho^{s-1},\\
\nonumber
\xi&=\alpha\rho^{s},\\
\label{eq3_37}
T&=\beta\rho^r.
\end{align}
where\[r\equiv\frac w{w+1},\] and parameters $\alpha,\beta>0$. In the case when the relaxation time $\tau=0$, we return to the case $\Pi=-3H\xi$ considered above.

Combining the Friedmann equation with the conservation equation, we obtain for the viscous pressure $\Pi$ and its time derivative $\dot{\Pi}$ the equations
\begin{align}
\nonumber
\Pi&=-\left[2\dot H+3H^2+w\rho\right],\\
\label{eq3_38}
\dot\Pi&=-\left[2\ddot H+6H\dot H+w\dot{\rho}\right].
\end{align}
Substituting these expressions into the evolution equation (\ref{eq3_36}) and limiting ourselves to the case of nonrelativistic matter ($w=0$, $r=0$), and assuming that the bulk viscosity is proportional to the Hubble parameter ($=1/2$) \cite{Chimento:1997vy}, we obtain
\begin{align}
\nonumber
\ddot H+b_1H\dot H-H^{-1}\dot H^2+b_2H^3=0,\\
\label{eq3_39}
b_1\equiv3\left(1+\frac1{\sqrt{3\alpha}}\right),\quad b_2\equiv\frac94\left(\frac2{\sqrt{3\alpha}}-1\right).
\end{align}
Using a number of simplifications, we came to a model with a single free parameter $\alpha$. Using known expressions for derivatives $\dot H$ and $\ddot H$, this parameter can be expressed through cosmographic parameters $q_0$ and $j_0$,
\begin{equation}\label{eq3_40}
\alpha=\sqrt3\frac{\frac12-q_0}{\frac{17}{4}-j_0+2q_0+q_0^2}.
\end{equation}
For the recent estimates of the cosmographic parameters already used above (\ref{recent_estimates}), we find $\alpha\approx0.6$ which perfectly agrees with the value of this parameter $\alpha\approx0.66$ found in the paper \cite{Mohan:2017poq} by complex calculations within the framework of the Israel-Stewart theory.
\subsection{Cosmological models with the creation of matter}
One of the interesting possibilities to avoid the introduction of dark energy(a component of an unknown nature) uses the fact that negative pressure---the key ingredient required to achieve accelerated expansion---naturally arises when the system deviates from thermodynamic equilibrium. In particular, as Zeldovich \cite{1970JETPLzeldovich} pointed out for the first time, negative pressure is generated in the process of particle creation due to the energy of the gravitational field. The problem to construct the models involving a substantially quantum process of particle production is related to the difficulty of including it in the classical Einstein's field equations. At the phenomenological level, these difficulties can be avoided \cite{Lima:2007kk}, choosing for a model an open thermodynamic system---a liquid with a non-conserved number of particles $N(t)$. In the expanding universe, the evolution of such a liquid will be described by the conservation equation
\begin{equation}
\label{eq3_41} \dot n+\vartheta n=n\Gamma.
\end{equation}
where $n\equiv N/V$ is the particle number density in the comoving volume $V$, $\vartheta=u_{;\mu}^\mu$ is the velocity of the expansion of the liquid ($\vartheta==u_{;\mu}^\mu=3H$ for the FLRW metric), and $\Gamma$ is the particle production rate in the comoving volume. $\Gamma>0$ corresponds to the particle creation, while $\Gamma<0$ describes the annihilation process.

In order to find the additional pressure generated by the particle production process, we express the first law of thermodynamics
\begin{equation}
\label{eq3_42} dE=dQ-pdV
\end{equation}
in terms of specific values: the energy density $\rho$ and the particle number density $n$. Here $dQ$ is the amount of heat that has entered the system over time $dt$. We introduce the energy density $\rho=E/V$, particle number density $n$, and specific heat $dq=dQ/N$. These quantities make it possible to transform (\ref{eq3_41}) to the form
\begin{equation}
\label{eq3_43} d\left(\frac\rho n\right)=dq-pd\left(\frac1n\right).
\end{equation}
We note that the conservation law (\ref{eq3_43}) is also valid in the case when the number of particles in the system is not conserved, i.e. $N=N(t)$.

Consider the spatially flat FLRW model of the Universe, interpreting it as an open thermodynamic system with the energy-momentum tensor
\begin{equation}
\label{eq3_44} T_{\mu\nu}=\left(\rho+p+\Pi\right)u_\mu u_\nu+\left(p+\Pi\right)g_{\mu\nu},
\end{equation}
where $\Pi$ is the additional pressure generated by the process of particle production due to the energy of the gravitational field.

Using the standard procedure for the transition from Einstein's field equations to the Friedmann equations, we obtain
\begin{align}
\nonumber \rho&=3H^2,\\
\label{eq3_45} \dot H&=-\frac12\left(\rho+p+\Pi\right),\quad8\pi G=1.
\end{align}
The conservation equation in this case takes the form
\begin{equation}
\label{eq3_46} \dot\rho+3H\left(\rho+p+\Pi\right)=0.
\end{equation}
We shall now show that for adiabatic processes the additional pressure $\Pi$ is determined by the rate of particle production $\Gamma$. To this end, we represent equation (\ref{eq3_42}) in the form
\begin{equation}
\label{eq3_47} Tds=d\left(\frac\rho n\right)+pd\left(\frac1n\right).
\end{equation}
Here $s\equiv S/N$ is the specific entropy. Using the conservation equations (\ref{eq3_41}) and (\ref{eq3_46}) for the derivatives $\dot n$ and $\dot\rho$ respectively, we obtain
\begin{equation}
\label{eq3_48} nT\dot s=-3H\Pi-\Gamma(\rho+p).
\end{equation}
For adiabatic processes $\dot s=0$ and, consequently,
\begin{equation}
\label{eq3_49} \Pi=-\frac{\Gamma}{3H}(\rho+p).
\end{equation}
Thus, we can state that, at least for adiabatic processes, the additional pressure due to the creation of particles is completely determined by the rate of this process. Interpreting $\Pi$ as the pressure generated by dissipative processes in a liquid (for example, bulk viscosity), we can say that a dissipative liquid is equivalent to an ideal fluid with a variable number of particles.

The effective equation of state parameter for the cosmological model with the creation of particles reads
\begin{align}
\nonumber w_{eff}=\frac{p_{tot}}{\rho_{tot}}=\frac{p+\Pi}{\rho},\\
\label{eq3_50} \Pi=-\frac{\Gamma}{3H}(\rho+p)=-\gamma\rho\frac\Gamma\vartheta,\quad\gamma\equiv w+1,\\
\nonumber w_{eff}=-1+\gamma\left(1-\frac\Gamma{3H}\right).
\end{align}
For $\Gamma<3H$ the parameter $w_{eff}>-1$, which corresponds to the quintessence, and for $\Gamma>3H$---to the phantom energy (with $w_{eff}<-1$). The case $\Gamma=3H$ realizes the cosmological constant.

Now let us calculate the main cosmological parameter (the deceleration parameter) in the model with particle creation. Using the relation \[q=-1-\frac{\dot H}{H^2}\] and
\begin{equation}
\label{eq3_51} \dot H=-\frac32\gamma H^2\left(1-\frac{\Gamma}{3H}\right),
\end{equation}
we find
\begin{equation}
\label{eq3_52} q=1+\frac32\gamma\left(1-\frac{\Gamma}{3H}\right).
\end{equation}
We now obtain the equation for the scale factor. Combining the modified Friedmann equations (\ref{eq3_45}) with the equation of state $p=w\rho$, we find the desired equation
\begin{equation}
\label{eq3_53} \frac{\ddot a}{a}+\frac{H^2}{2}\left[1+3w-\frac{1+w}{H}\Gamma\right].
\end{equation}
Setting $\Gamma=0$, we reproduce the standard second Friedmann equation.

Models with the creation of matter ($\Gamma\ne0$) allow us to describe the evolution of the Universe, beginning with the postinflationary phase and ending with the current stage of accelerated expansion \cite{Lima:2007kk,Pan:2013rha,Pan:2016jli,Chakraborty:2014fia,Lima:2015xpa,Ramos:2014dba,Chakraborty:2014ora}. Moreover, within the framework of such models, it is possible to reproduce a return in the future to the decelerated expansion, interpreting the observed accelerated expansion as a transient process \cite{Pan:2013rha,Chakraborty:2014ora}. We emphasize that the attractive feature of models with the creation of matter is the absence of any exotic component like the dark energy.

In attempt to describe the complete evolution of the Universe, we should consider the rate of matter creation as a function $\Gamma(H)$ of the expansion rate---the Hubble parameter $H$. Various functions of the form \[\Gamma(H)=\sum\limits_i\Gamma_iH^i\] have been studied recently [cf. \cite{Chakraborty:2014fia}].

How adequately do these models describe the temporal evolution of the Universe? In a model using the matter creation rate as a power series in the Hubble parameter, a set of additional parameters $\Gamma_i$ arises. Therefore, any attempt to apply the model should be preceded by the determination of the values of these parameters. A natural question arises: should we make additional cosmological measurements to determine these parameters, or they can be expressed through already known quantities? Below we give the positive answer to the question posed. We will show that the parameters that determine the particle creation rate can be analytically expressed through the cosmographic parameters, using the approach proposed in the present paper.

The relation (\ref{eq3_51}) rewritten in the form
\begin{equation}
\label{eq3_54} \Gamma=3\left(1+\frac{2}{3(1+w)}\frac{\dot H}{H^2}\right)
\end{equation}
determines the time dependence of the matter creation rate if we know the evolution of the Universe $H(z)$. The function $H(z)$ can only be calculated within the framework of a certain model, for example, within the framework of the Standard Cosmological Model, in which, however, there is no particle production process. Inclusion of this process in the model will inevitably require {\it ad hoc} specification of the function $\Gamma(H)$. The simplest way out of the situation is the phenomenological construction of the function $\Gamma(H)$, relying on available information on known stages of the evolution of the Universe.

To achieve this goal, we break the history of the Universe into three stages: the early radiation-dominating Universe, the stage of slow expansion with the dominance of matter and the current stage of accelerated expansion. Back in 1998, Gunzing et al. \cite{Gunzig:1997tk} formulated a number of requirements that the function $\Gamma(H)$ must satisfy in order not to conflict with the firmly established features of the early stage of the evolution of the Universe. It turned out that the only function that satisfies these requirements is the rate of creation of matter proportional to the energy density, i.e. $\Gamma(H\propto H^2)$. For the intermediate phase with the dominance of matter, the simplest choice is $\Gamma(H)\propto H$. With this choice of the rate of matter creation, equation (\ref{eq3_51}) is easily solved, leading to $H\propto t^{-1}$ and the natural power-law dependence of the scale factor on time. With the growth of time (decrease in density), the mechanism should start working that generates the accelerated expansion of the Universe due to the negative pressure generated by the process of particle creation. As is known, the current stage of the evolution of the Universe is well described by the standard cosmological model. The evolution equation for the Hubble parameter in this model reads
\begin{equation}
\label{eq3_55} \dot H+\frac32H^2\left[1-\left(\frac{H_f}{H}\right)^2\right]=0,
\end{equation}
where $H_f=\sqrt{\Lambda/3}$ is the de Sitter asymptotics of the Hubble parameter ($H\ge H_f$), $\Lambda$ is the cosmological constant. Therefore, the explicit form of the function $\Gamma(H)$ can be found from the requirement that the evolution equation for the model with the creation of matter (\ref{eq3_51}) leads to the cosmological evolution that coincides with the one in the Standard Model. Comparing (\ref{eq3_51}) and (\ref{eq3_55}), we find
\begin{equation}
\label{eq3_56} \frac{\Gamma}{3H}=\left(\frac{H_f}{H}\right),
\end{equation}
or $\Gamma\propto H^{-1}$. The limiting value $\Gamma=3H_f$ corresponds strictly to the de Sitter phase: $\dot H = 0$, $H=H_f=const$. Finally, the constant matter creation rate $\Gamma=\mathrm{const}$ allows reproduction of cosmological models without initial singularities \cite{Chakraborty:2014fia}. Various linear combinations of the terms $\Gamma=\mathrm{const}$, $\Gamma\propto H$, $\Gamma\propto H^2$, and $\Gamma\propto H^{-1}$ allow us to construct a wide class of prospective cosmological models.

Let us now consider the procedure for determination of the matter creation rate parameters for various types of the $\Gamma(H)$ dependencies. We begin with the case
\begin{equation}
\label{eq3_57} \Gamma(H)=\Gamma_0+\Gamma_{-1}/H,
\end{equation}
where the model parameters $\Gamma_0$ and $\Gamma_1$ are constant values, which must be expressed in terms of the current values of the cosmographic parameters.

The system of equations for determination of the required parameters reads
\begin{align}
\nonumber \frac{\dot H}{\alpha H^2}&=1-\frac13\frac{\Gamma_0}{H}-\frac13\frac{\Gamma_{-1}}{H^2},\\
\label{eq3_58} \frac{\ddot H}{\alpha\dot H H}&=2-\frac13\frac{\Gamma_0}{H},\quad \alpha\equiv-\frac32(1+w).
\end{align}
The left-hand sides of the system (\ref{eq3_58}) represent dimensionless combinations of the cosmographic parameters. Using the expressions for the time derivatives of the Hubble parameter, we find
\begin{align}
\nonumber \frac{\dot H}{\alpha H^2}&=\frac1\alpha(1+q),\\
\label{eq3_59} \frac{\ddot H}{\alpha\dot H H}&\equiv f_1=\frac23\frac{(j+3q+2)}{(1+w)(1+q)}.
\end{align}
Solutions of the system (\ref{eq3_58}) in terms of the cosmographic parameters read
\begin{align}
\nonumber \frac{\Gamma_0}{H}&=3(2-f_1)=2\frac{-j+3w(1+q)+1}{(1+q)(1+w)},\\
\label{eq3_60} \frac{\Gamma_{-1}}{H^2}&=3\left(f_1-1-\frac23\frac{1+q}{1+w}\right)=-\frac{-2j+2q^2+3(1+q)+q+1}{(1+w)(1+q)}.
\end{align}
We recall that the right-hand sides of the relations (\ref{eq3_60}) must be calculated at the current values of the cosmographic parameters.

The solutions obtained above can be tested as follows. The deceleration parameter $q$ in the model with particle creation is
\begin{equation}
\label{eq3_61} q=-\frac{\ddot a}{aH^2}=-1+\frac32(1+w)\left(1-\frac{\Gamma}{3H}\right).
\end{equation}
Hence we find
\begin{equation}
\label{eq3_62} \Gamma=H\frac{1-2q+3w}{1+w}.
\end{equation}
The solutions (\ref{eq3_60}) found above are, generally speaking, valid for any instance of time. Therefore, when substituting them in (\ref{eq3_57}), we must reproduce the relationship between the deceleration parameter and the particle creation rate (\ref{eq3_62}). Performing this substitution, we see that for the solutions found, the relation (\ref{eq3_62}) is satisfied.
Using a similar procedure, we consider a number of models below.

A model including the constant and linear terms
\begin{equation}
\label{eq3_63} \Gamma=\Gamma_0+\Gamma_1H.
\end{equation}
similar to the former case, contains two free parameters, so to find them it suffices to consider the first and the second time derivatives of the Hubble parameter. The solutions obtained are:
\begin{align}
\nonumber \frac{\Gamma_0}{H}&=2\frac{j-q(1+2q)}{(1+q)(1+w)},\\
\label{eq3_64} \Gamma_1&=\frac{-2j+2q^2+3w(1+q)+q+1}{(1+w)(1+q)}.
\end{align}
It is easy to verify that the above-found values of the parameters satisfy the test relation (\ref{eq3_62}).

Parameters for the model
\begin{equation}
\label{eq3_65} \Gamma=\Gamma_0+\Gamma_2H^2.
\end{equation}
are equal
\begin{align}
\nonumber \frac{\Gamma_0}{H}&=\frac{2j+3q(-2q+w-1)+3w+1}{2(1+q)(1+w)},\\
\label{eq3_66} \Gamma_2H&=\frac{-2j+2q^2+3w(1+q)+q+1}{2(1+w)(1+q)}.
\end{align}

In order to find all the free parameters of the model
\begin{equation}
\label{eq3_67} \Gamma=\Gamma_0+\Gamma_1H+\Gamma_{-1}/H.
\end{equation}
we must include into consideration the third-order time derivative $\dddot H$. The system of equations for determination of the parameters then reads
\begin{align}
\nonumber \frac{\dot H}{\alpha H^2}&=1-\frac13\left(\frac{\Gamma_0}{H}+\Gamma_1+\frac{\Gamma_{-1}}{H^2}\right),\\
\label{eq3_68} \frac{\ddot H}{\alpha\dot H H}&=2-\frac23\Gamma_1-\frac13\frac{\Gamma_0}{H},\\
\nonumber \frac{\dddot H}{\alpha\ddot H H}&=\left(2-\frac23\Gamma_1\right)\frac{\dot H^2}{\ddot H H}+\left(2-\frac13\frac{\Gamma_0}{H}-\frac23\Gamma_1\right).
\end{align}
All the constructions \(\dot H/(H^2)\), \(\ddot H/(\dot H H)\), \(\dddot H/(\ddot H H)\), and \(\dot H^2/(\ddot H H)\), in the system of equations (\ref{eq3_68}) are known dimensionless functions of the cosmographic parameters. Solving the system, we find for the free parameters
\begin{align}
\nonumber \frac{\Gamma_0}{H}&=-\frac{2\left[j^2+j(q(q+4)+1)+q(2q+s+1)+s\right]}{(1+q)^3(1+w)},\\
\label{eq3_69} \Gamma_1&=\frac{j^2+2jq+3q^2+qs+3w(1+q)^3+3q+s+1}{(1+q)^3(1+w)},\\
\nonumber \frac{\Gamma_{-1}}{H^2}&=\frac{j^2+2j(q(q+3)+1)+q(s-q(q+2)(2q+1)+s}{(1+q)^3(1+w)}.
\end{align}
Finally, in the most general of the models we are considering, the matter creation rate has the form
\begin{equation}
\label{eq3_70} \Gamma=\Gamma_0+\Gamma_1H+\Gamma_2H^2+\Gamma_{-1}/H.
\end{equation}

$$ \Gamma_0 =\frac{-2 H(t) }{(q+1)^2 (w+1)
   \left(3 j^2+6 j \left(q^2+5 q+3\right)+2 q^4+26 q^3+87 q^2+86 q+26\right)}$$
   $$\bigl(2 j^3 (3 q+5)+j^2 \left(8 q^3+63 q^2+104 q-s+38\right)+j
   \bigl(-l (q+1)+2 q^5+34 q^4+$$
   $$+174 q^3+2 q^2 (s+145)+q (3 s+164)+3 s+26\bigr)-l \left(q^3+6 q^2+8
   q+3\right)+10 q^5+$$
   $$+2 q^4 s+77 q^4+19 q^3 s+156 q^3+49 q^2 s+112 q^2+50 q s+26 q+17 s\bigr)$$
   $$\Gamma_1= \frac{1}{(q+1)^2 (w+1) \left(3 j^2+6 j \left(q^2+5 q+3\right)+2 q^4+26 q^3+87 q^2+86
   q+26\right)}$$
   $$\biggl(2
   j^3 (3 q+5)+j^2 \left(2 q^3+q^2 (9 w+38)+18 q (w+4)-s+9 w+25\right)+j \bigl(-l (q+1)+$$
   $$+2 q^4 (9 w+2)+6 q^3
   (21 w+13)+2 q^2 (3 s+126 w+86)+q (11 s+198 w+106)+7 s+54 w+$$
   $$+18\bigr)-l \left(3 q^2+5 q+2\right)+6 q^6 w+90
   q^5 w+6 q^5+2 q^4 s+423 q^4 w+96 q^4+26 q^3 s+858 q^3 w+$$
   $$+266 q^3+66 q^2 s+855 q^2 w+285 q^2+63 q s+414 q
   w+138 q+20 s+78 w+26\biggr)$$
   $$\Gamma_{-1}= \frac{H(t)^2}{3 (q+1)^2 (w+1) \left(3 j^2+6 j \left(q^2+5 q+3\right)+2 q^4+26 q^3+87 q^2+86
   q+26\right)}$$
   $$ \biggl(12 j^3 (2 q+3)+j^2 \left(24 q^3+259 q^2+434 q-3
   s+166\right)+j \bigl(-3 l (q+1)-24 q^5-46 q^4+$$
   $$+454 q^3+2 q^2 (s+559)+q (s+768)+5 s+156\bigr)-l \left(4
   q^3+21 q^2+27 q+10\right)-12 q^7-$$
   $$-174 q^6-726 q^5+6 q^4 s-1137 q^4+50 q^3 s-720 q^3+130 q^2 s-156 q^2+137 q
   s+48 s\biggr)$$
   $$\Gamma_2= -\frac{2}{3 (q+1) (w+1) H(t) \left(3 j^2+6 j \left(q^2+5
   q+3\right)+2 q^4+26 q^3+87 q^2+86 q+26\right)}$$
   $$ \bigl(3 j^3+j^2 (11 q+2)+j \left(-2 q^3+13 q^2+q (4 s+6)+4
   s\right)+l (q+1)^2-$$
   $$ -6 q^4-3 q^3+7 q^2 s+10 q s+3 s\bigr)$$
   The above ratios for the parameters will allow, in particular, to answer the question\cite{Lima:2007kk}: will there be again a transition from acceleration to deceleration in the framework of the creation model?

\subsection{Cosmography of Cardassian Model}

In this section, the focus of our attention will be the so-called Cardassian model (CM) \cite{Freese:2002sq,Gondolo:2002fh}, which allows us to describe the accelerated expansion of the Universe filled with "habitual" components: matter and radiation.

As an alternative explanation for the observed accelerated expansion of the Universe, Freese and Lewis \cite{Freese:2002sq} proposed a modified version of the first Friedman equation
\begin{equation}
\label{cm_1} H^2=g(\rho_m),
\end{equation}
where the energy density of the flat Universe $\rho$ includes only non-relativistic matter (both baryonic and dark) and radiation, but does not contain dark energy. We will work further with the simplest version of the CM, which uses an additional power law term on the right-hand side of the Friedmann equation
\begin{equation}
\label{cm_2} H^2=A\rho_m+B\rho_m^n.
\end{equation}
Note that $B=0$ in the standard FLRW cosmology. Therefore, we must choose $A=8\pi G/3$.

Suppose that the Universe is filled only with non-relativistic matter. In this case, with the dominance of the second term (high densities, late Universe): $H\propto\rho_m^{n/2}\propto a^{-3n/2}$, $\dot a\propto a^{-3n/2+1}$, $a\propto t^{2/(3n)}$. Therefore, the expansion is accelerated for $n<2/3$. On the upper border for $n=2/3$ we have $a\propto t$ and $\ddot a=0$; for $n=1/3$ we have $a\propto t^2$ (the acceleration is constant). For $n>1/3$ the acceleration is diminishing in time, while for $n<1/3$ the acceleration is increasing. It is interesting to note that if $n=2/3$ we have $H^2\propto a^{-2}$: in a flat Universe, the term similar to the curvature term is generated by matter.

Let us represent the energy density of the Cardassian model in the form of a sum of densities of ordinary matter $\rho_m$ and a component with density $\rho_X=\rho_m^n$ so that $H^2\propto \rho_m+\rho_X$. As we have seen above, $a\propto t^{2/(3n)}$ in the case of dominance of the additional term in the Friedmann equation. Since \[a\propto t^{\frac2{3(w+1)}},\] we find that the parameter of the equation of state is
\begin{equation}
\label{cm_3} w_X=n-1.
\end{equation}
This relation holds for any arbitrary one-component liquid with $w_X=\mathrm{const}$. In this case
\begin{equation}
\label{cm_4} \frac{d\rho_X}{dz}=3\rho_X\frac{1+w_X(z)}{1+z}.
\end{equation}
Taking into account that \[\rho_X=\rho_m^n=(1+z)^{3(w_X+1)n}\], after substitution of \begin{equation}
\label{cm_5} w_x=n-1.
\end{equation}
 into (\ref{cm_4}) for $0<n<2/3$ we find
\begin{equation}
\label{cm_6} -1<w_x<-1/3.
\end{equation}
As expected, this interval of the parameter $w_X$ values generates negative pressure, responsible for the late time accelerated expansion of the Universe.

We now calculate the deceleration parameter in the CM. Let us now apply the cosmographic approach to find the parameters $(B,n)$ for the CM. The evolution of the CM is described by the system of equations
\begin{align}
\label{cm_7} H^2&=A\rho+B\rho^n;\\
\label{cm_8} \dot\rho&+3H\rho=0.
\end{align}
Differentiating the equation (\ref{cm_7}) with respect to cosmological time and using (\ref{cm_8}) we transform the system to the form
\begin{align}
\label{cm_9} H^2&=A\rho_m+B\rho^n_m;\\
\label{cm_10} -\frac23\dot H&=A\rho_m+Bn\rho^n_m.
\end{align}
Solutions of this system read
\begin{align}
\label{cm_11} \rho_m&=-\frac{nH^2+\frac23\dot H}{A(1-n)};\\
\label{cm_12} B&=\frac{H^2+\frac23\dot H}{\rho^n(1-n)}.
\end{align}
To calculate the parameter $n$, we need an expression for $\ddot H$,
\begin{equation}
\label{cm_13} \frac29\frac{\ddot H}{H}=A\rho_m+n^2B\rho_m^n.
\end{equation}
Substituting the above solutions (\ref{cm_11}) and (\ref{cm_12}) for $\rho$ and $B$ into this expression, we obtain
\begin{equation}
\label{cm_14} \frac29\frac{\ddot H}{H}=-n+\frac23\frac{\dot H}{H^2}(1+n).
\end{equation}
Hence for the parameter $n$ we find
\begin{equation}
\label{cm_15} n=-\frac{\frac23\left(\frac13\frac{\ddot H}{H^3}+\frac{\dot H}{H^2}\right)}{1+\frac23\frac{\dot H}{H}}.
\end{equation}
Expressions (\ref{cm_4}) and (\ref{cm_7}) allow us to express the parameters of CM through the cosmographic parameters. Using the known expressions for the time derivatives of the Hubble parameter in terms of the cosmological parameters (**), we find
\begin{align}
\label{cm_16} \frac{B\rho_m^n}{H^2}=\frac13(1-2q);\\
\label{cm_17} n=\frac23\frac{j-1}{2q-1}.
\end{align}
Note that the parameters $B$ and $n$ are constants, which was explicitly used in deriving the relations obtained above. Let us verify that the solutions (\ref{cm_16}) and (\ref{cm_17}) found above agree with this condition. The requirement $\dot B=0$ is transformed into (\ref{cm_14}) and, consequently, it is consistent with the above obtained expression for $n$. The constancy of the parameters allows us to calculate them for the values of the cosmological parameters at any time instance. Since the main body of information about cosmological parameters refers to the current time $t_0$, the relations (\ref{cm_16}) and (\ref{cm_17}) can be represented in the form
\begin{align}
\label{cm_18} \frac{B\rho_0^n}{H_0^2}=\frac13(1-2q_0);\\
\nonumber n=\frac23\frac{j_0-1}{2q_0-1}.
\end{align}
We should interpret the time-dependent solution for the density differently. Using (\ref{cm_16}), it can be represented in the form
\begin{equation}
\label{cm_20} \frac\rho{\rho_c}=\frac{-n+\frac23(1+q)}{1-n},\quad\rho_c\equiv\frac{3H^2}{8\pi G}.
\end{equation}
The current density value in CM can be found by the substitution $q\to q_0$, $H\to H_0$.

It is interesting to note that the above-found expression (\ref{cm_2}) for the parameter $n$ coincides exactly with the parameter $s$, which is one of the so-called state finders\cite{Sahni:2002fz}
\begin{equation}
\label{cm_21} r\equiv\frac{\dot a}{aH^3},\quad s=\frac23\frac{r-1}{2q-1}.
\end{equation}
The coincidence is obvious, since $r\equiv j$. The reason for the coincidence can be explained as follows. In any model for which the scale factor $a\propto t^\alpha$ the simple relations hold for the cosmographic parameters $q$ and $j$
\begin{equation}
\label{cm_22} 2q-1=\frac{2-3\alpha}{\alpha},\quad j-1=\frac{2-3\alpha}{\alpha^2}.
\end{equation}
In the Cardassian model \[a\propto t^{\frac2{3n}},\] which implies that $s=n$.

Using the expression (\ref{cm_9}) for $\dddot H$ and the above-found solutions (\ref{cm_11}, \ref{cm_12}, \ref{cm_17}), we obtain an equation relating the cosmological parameters
\begin{align}
\label{cm_23} s&+qj+(3n+2)j-2q(3n-1)=0;\\
\nonumber n&=\frac23\frac{j-1}{2q-1}.
\end{align}
This is a fourth-order ODE for the scale factor. For $n=0$ the equation (\ref{cm_15}) reproduces the known \cite{Dunajski:2008tg} relation between the cosmographic parameters in the LCDM
\begin{equation}
\label{cm_24} s+2(q+j)+qj=0.
\end{equation}

Let us now turn to the dimensionless form of the evolution equation (\ref{cm_7}) for the scale factor $a(t)$, the coefficients of which are expressed in terms of the cosmological parameters. To this end, using \[\rho=\frac{\rho_0}{a^3},\] the Friedmann equation can be rewritten in the form
\begin{equation}
\label{cm_25} \dot a^2=A\rho_0 a^{-1}+B\rho_0^na^{-3n+2}.
\end{equation}
Transferring to the dimensionless time $\tau=H_0t$ and substituting in (\ref{cm_25}), we obtain
\begin{equation}
\label{cm_26} \rho_0=-\frac{nH_0^2+\frac23 \dot H_0}{A(1-n)},\quad B=\frac{H_0^2+\frac23\dot H_0}{\rho_0^n(1-n)}.
\end{equation}
we transform the Friedmann equation (\ref{cm_25}) to the form
\begin{align}
\nonumber \frac{da}{d\tau}&=\left[F(q_0,j_0)a^{-1}+\Phi(q_0,j_0)a^{-3n+2}\right]^{1/2},\\
\nonumber F&=-\frac{n-\frac23(1+q_0)}{1-n},\quad\Phi=\frac{1-\frac23(1+q_0)}{1-n},\\
\label{cm_27} n&=-\frac23\frac{j_0-1}{2q_0-1}.
\end{align}
This equation should be solved with the constraints $n<2/3$ (a condition ensuring the accelerated expansion of the Universe) and $n<\frac23(1+q_0)$ (a condition ensuring the positivity of the energy density). It is easy to see that the two conditions are consistent, since in the case of accelerated expansion $q<0$.
\begin{figure}
\centerline{\psfig{file=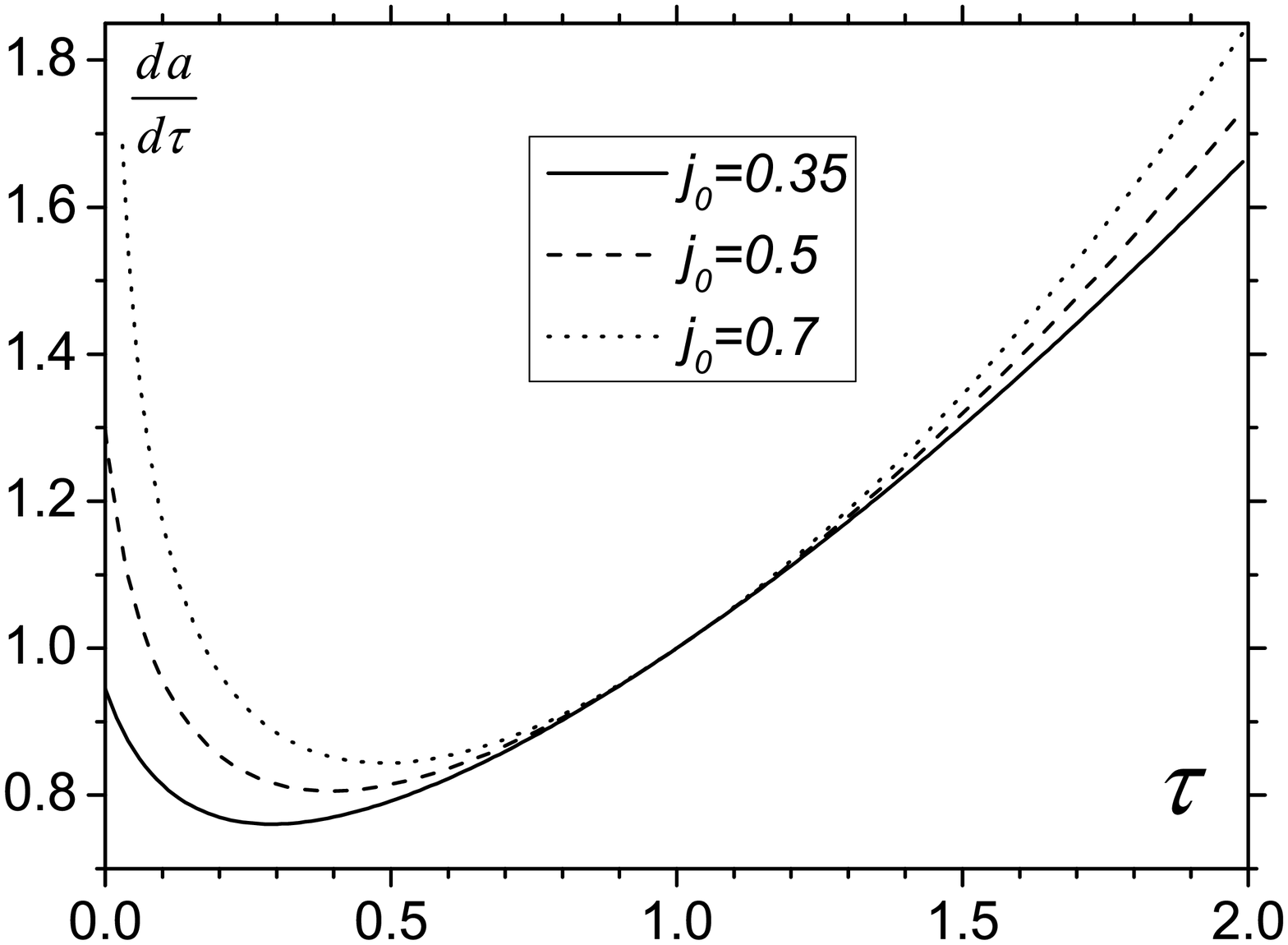,width=0.48\textwidth}
\psfig{file=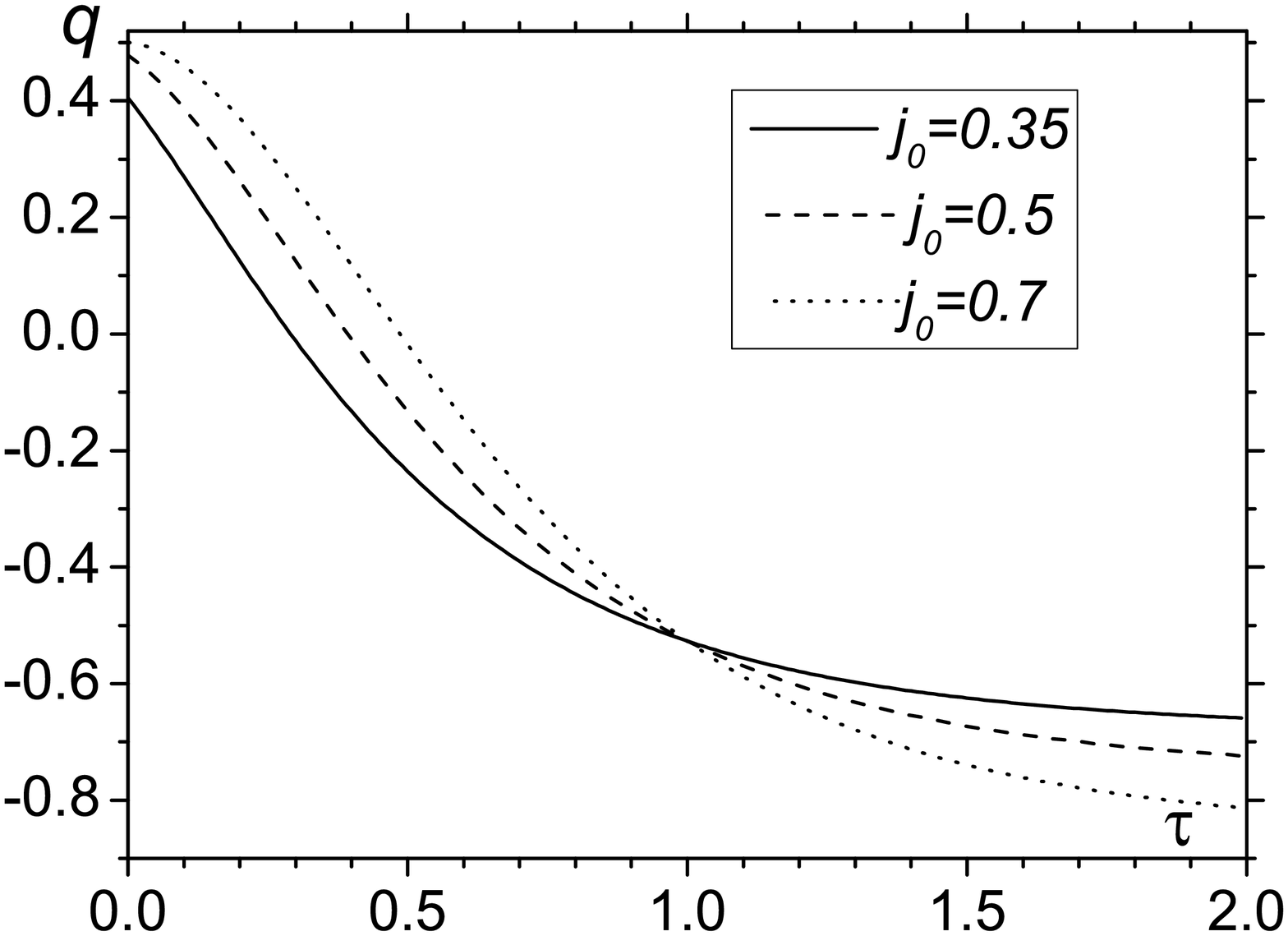,width=0.48\textwidth}}

\caption{Demonstration of the transition from the delayed expansion to the accelerated one for different values of the parameter $j$ ($q_0=-0.527$ for all the three options). Left figure: derivative $da/d\tau$ as a function
of dimensionless time. Right figure: the deceleration parameter $q$ as a function of dimensionless time. \label{a1q}}

\end{figure}
Fig. \ref{a1q} illustrates the transition from the delayed expansion to the accelerated one for the late-time Universe in the CM. Growth of the parameter $j_0$ leads to a decrease in the index $n$: $j=0.35\Rightarrow n\approx0.21$, $j=0.5\Rightarrow n\approx0.16$, $j=0.7\Rightarrow n\approx0.097$. At the limit $j\to1$ (LDCM), as expected, the parameter $n\to0$.

\begin{figure}

\centerline{\psfig{file=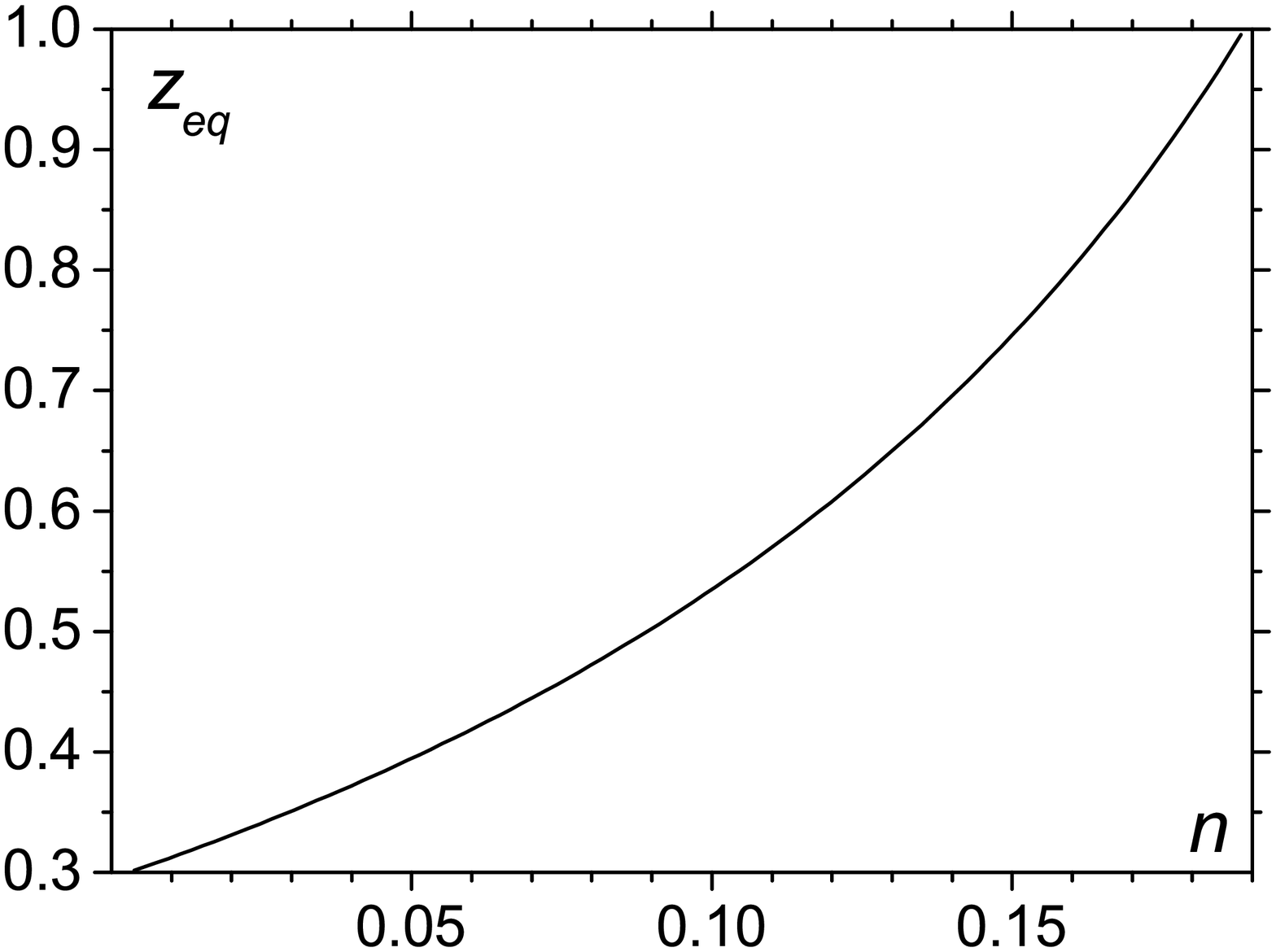,width=6.7cm}}
\vspace*{8pt}
\caption{\label{z}Function $z_{eq}(n,q_0)$ at the value of the current deceleration parameter $q_0=-0.527$.}
\end{figure}

The authors of the CM suggested the following procedure to estimate the parameter $B$ \cite{Freese:2002sq}. The initial CM is described by the set of parameters $\left\{B,n\right\}$. We make the transition $\left\{B,n\right\}\to\left\{z_{eq},n\right\}$ to a new set of parameters, where $z_{eq}$ is the redshift value, at which the contributions of the terms $A\rho_m$ and $B\rho_m^n$ are equal:
\begin{equation}
\label{cm_28} A\rho(z_{eq})=B\rho^n(z_{eq}).
\end{equation}
Since $\rho=\rho_0/a^3=\rho_0(1+z)^3$, then
\begin{equation}
\label{cm_29} \frac B A=\rho_0^{1-n}(1+z_{eq})^{3(1-n)}.
\end{equation}
Using the expression
\begin{equation}
\label{cm_30} A=\frac{H_0^2}{\rho_0}-B\rho_0^{n-1}.
\end{equation}
for the parameter $A$, we find
\begin{equation}
\label{cm_31} \frac{B\rho_0^n}{H_0^2}=\frac1{1+(1+z_{eq})^{3(n-1)}}.
\end{equation}
According to the authors of the model to match the CMB and the supernovae data, the value $z_{eq}$ should lie in the range $0.3<z_{eq}<1$, although a more thorough analysis, in principle, allows to narrow this interval. Comparing (\ref{cm_18}) and (\ref{cm_31}), one can see the obvious advantage of the cosmographic approach: to find the parameter $B$, we did not have to introduce additional parameters. The dimensionless parameter $B\rho_0^n/H_0^2$ is determined by the current value of the fundamental cosmological parameter---the deceleration parameter $q_0$. Equating (\ref{cm_18}) and (\ref{cm_31}), we find the function $z_{eq}(n,q_0)$ that allows us to estimate the interval of variation of the parameter $n$ corresponding to the range $0.3<z_{eq}<1$. We see (cf. Fig. \ref{z}) that this interval includes the parameter values $n\le0.15$.

 The method presents an interesting possibility to calculate the highest order cosmological parameters from the values of the lowest parameters, which are known with better accuracy. For example, formula (\ref{cm_23}) can be used to estimate the parameter $s_0$ for known values $q_0$ and $j_0$,
\begin{align}
\nonumber s_0&=-q_0j_0-(3n+2)j_0+2q_0(3n-1),\\
\label{cm_32} n&=\frac23\frac{j_0-1}{2q_0-1}.
\end{align}
In particular, $j=1$ in the LCDM and the relation (\ref{cm_32}) is transformed into $s=-3q-2$. It is easy to see that the cosmographic parameters of the LCDM
\begin{align}
\nonumber q&=-1+\frac32\Omega_m,\\
\label{cm_33} s&=1-\frac92\Omega_m.
\end{align}
exactly satisfy this relation.

\section{\label{analysis}Cosmographic analysis of models with interactions of the dark components}
\subsection{Dark sector interaction}
Since the nature of the dark components of the Universe (dark matter and dark energy unknown, it is possible that these components can interact with each other. At the phenomenological level, the interaction between dark energy with density
$\rho _{de}$
 and cold dark matter with density $\rho _{dm}$
can be described by a system of equations\cite{Amendola:1999er,Zimdahl:2001ar,Bolotin:2013jpa}
	
\begin{align}
    \nonumber \dot{\rho} _{dm}  + 3H \rho _{dm} &= Q \hfill  \\
  \label{eq7.1} \dot {\rho}_{de}  + 3H( \rho _{de}+ p_{de} ) &= - Q
\end{align}
 The function of $Q$ s known  as the interaction function, generally speaking, depends on scale factor. The sign of $Q$
 defines the direction of the flux of energy.

The system  (\ref{eq7.1}) can be given an alternative interpretation. It is convenient to introduce the effective pressures
$\Pi _{dm} $ and $\Pi _{de} $ \cite{Zimdahl:2001ar}.
\begin{equation}\label{eq7.2}
 Q \equiv  - 3H\Pi _{dm}  =  + 3H\Pi _{de}
\end{equation}

The introduction of effective pressure allows you to go from (\ref{eq7.1})to the system of equations.
  \begin{align}
   \nonumber  \dot \rho _{dm}  + 3H(\rho _{dm}  + \Pi _{dm} ) &= 0 \hfill  \\
   \label{eq7.3}\dot \rho _{de}  + 3H\left( {\rho _{de}  + p_{de}  + \Pi _{de} } \right) &= 0
\end{align}

In this case, the conservation equation system formally look as those for two independent fluids.
A coupling between them has been mapped into the relation $\Pi _{dm}  =  - \Pi _{de}. $
In general, the coupling term $Q$  can take any possible form $Q = Q\left( {H,\rho _{dm} ,\rho _{de} ,t} \right).$
However, physically, it makes more sense that the coupling be time-independent. Among the time-independent options, preference is given to a factorized $H$  dependence
$Q = H \, q(\rho _{dm} ,\rho _{de} ).$ During this kind of factorization, the effects of the coupling on the dynamics of $\rho _{dm} $ and $\rho _{de} $
become effectively independent from the evolution of the Hubble scale $H.$ The latter is related to the fact that the time derivatives  that go into the conservation equation can be transformed in the following way $d/dt \to H\,d/d\,lna. $  . It is important to note\cite{Wands:2012vg},  that the decoupling of the dynamics of the two dark components from $H$
 is valid in any theory of gravity, because it is based on the conservation equations. Any coupling of this type can be approximated at late times by a linear expansion
 \begin{equation}\label{eq7.4}
 Q = q_0^*  + q_{dm}^* \left( {\rho _{dm}  - \rho _{dm,0} } \right) + q_{de}^* \left( {\rho _{de}  - \rho _{de,0} } \right).
 \end{equation}
 Constants $ q_0^* ,q_{dm}^* ,q_{de}^* $ can always be redefined in order to put coupling $Q$  in the form
  \begin{equation}\label{eq7.5}
 Q = q_0  + q_{dm} \rho _{dm}  + q_{de} \rho _{de}
  \end{equation}
  Special cases
  \begin{align}
  \nonumber Q \propto \rho _{dm} & ,\quad q_0  = q_{de}  = 0; \hfill  \\
 \label{eq7.6}  Q \propto \rho _{de} &,\quad q_0  = q_{dm}  = 0; \hfill  \\
 \nonumber Q \propto \rho _{total,\quad } & q_0  = 0,\;q_{dm}  = q_{de}.
\end{align}
Previously, we analysed linear interactions. However, from a physical point of view, an interaction between two components should depend on the product of the abundances of the individual components, as, for instance,  in chemical or nuclear reactions. Consequently, a product coupling, i.e., an interaction proportional to the product of dark matter and dark energy densities looks more appealing. Analysis of cosmological models with specific non-linear interactions was performed in\cite{Arevalo:2011hh}.  The authors considered nonlinear interaction of the form
\begin{equation}\label{eq7.7}
Q \propto H\rho _{de}^{m - n} \rho _{dm}^n .
  \end{equation}
 The ansatz (1.17)  also includes the previously analysed linear cases. The combination
 $\left( {m,n,s} \right) = \left( {1,1, - 1} \right)$ corresponds to $Q \propto H\rho _{dm} $
 while $\left( {m,n,s} \right) = \left( {1,0, - 1} \right)$  reproduces $Q \propto H\rho _{de}. $
 \subsection{Cosmography of the dynamical cosmological constant}
 At the present time, the interpretation of the dark energy in the form of the cosmological constant $\Lambda $
  as energy of  physical vacuum is the most supported among other alternatives. It automatically leads to the equation of state of this  substance
  $p_\Lambda   =  - \rho _\Lambda  $
 ($p_\Lambda  $is the pressure, and  $\rho _\Lambda  $
 is the energy density), ensuring the accelerating expansion of the  Universe. The hypothesis allowing resolving a number of current cosmological problems involves moving to a time-dependent
$\Lambda  \to \Lambda (t).$
 In virtue of the energy conservation law, the vacuum decay should be accompanied by  changing the dark matter energy density $\rho _{_m }. $
 Dynamics of the two-component system can be described by the system of  equations comprising the first Friedmann equation and the conservation equation
\begin{align}
\nonumber \rho _m  + \rho _\Lambda   = 3H^2 , \\
\label{eq7.8} \dot \rho _m  + 3H\rho _m  =  - \dot \rho _\Lambda
 \end{align}
We took into account that for the cold dark matter the pressure $p_m  = 0..$ In the  right part of the conservation equation there is a new term $ - \dot \rho _\Lambda$
 playing the role of the source generated by the decaying CC. From the system (\ref{eq7.8}) one can obtain the equation for the Hubble parameter
 \begin{equation}\label{eq7.9}
 2\dot H + 3H^2  - \rho _\Lambda   = 0.
 \end{equation}

 At the phenomenological level to solve the equation (2) a model of the cosmological constant decay is needed. Below, we will consider a  simple, but fully analyzable model \cite{Carneiro:2007bf},
	 \begin{equation}\label{eq7.10}
\rho _\Lambda   = \sigma H
 \end{equation}
We will show that cosmography allows to easy express a single parameter of the model  in terms of the decoration parameter.
The two-parametric model suggested in the work \cite{Szydlowski:2015bwa} also treats the cosmological constant as the decaying vacuum energy. More specifically, the ideology of this model goes back to the hypothesis about existence of an unstable false vacuum\cite{Coleman:1977py,Callan:1977pt}. If $E_0 ^{(false)} $
and $E_0 ^{(true)} $
 are energies of the false and the true vacuums, the hypothesis involves a universal behavior of the substance initially having been in the false vacuum
\begin{equation}\label{eq7.11}
E_0 ^{(false)}  = E_0 ^{(true)}  + \frac{\alpha }{{t^2 }} \pm  \ldots for\quad t >  > T,
\end{equation}
where $T$is the typical time of the tunneling from the false vacuum to the true one. In terms of the time-depending the cosmological constant relation (\ref{eq7.11}) can be rewritten as\cite{Szydlowski:2015bwa}
\begin{equation}\label{eq7.12}
\rho _\Lambda  (t) = \Lambda (t) = \Lambda _{bare}  + \frac{\alpha }{{t^2 }}
\end{equation}
that emerges from the covariant theory of a scalar field and presents a leading term at the late time of evolution.
Now we apply our procedure for finding parameters to the two models under consideration. For the first considering model (\ref{eq7.10}) the evolution equation (\ref{eq7.9}) reads
\begin{equation}\label{eq7.13}
2\dot H + 3H^2  - \sigma H = 0.
\end{equation}

Using relation $
\dot H $
we express the free parameter of the model$\sigma$ in a simple way through the Hubble parameter and the deceleration parameter:
\begin{equation}\label{eq7.14}
\sigma  = H_0 \left( {1 - 2q_0 } \right).
\end{equation}

The second considering model (\ref{eq7.12}) may be transform to the new parameterization of the vacuum dark energy using $H \propto 1/t$ [7]
\begin{equation}\label{eq7.15}
\Lambda (H) = \Lambda _0  + 3\beta H^2 ,
\end{equation}
where there are two free parameters$
\Lambda _0  \equiv  \equiv \Lambda _{bare} $    and $\beta.$
Hereby, the evolution equation (\ref{eq7.9}) takes the form
\begin{equation}\label{eq7.16}
\dot H = \frac{{\Lambda _0 }}{2} - \delta H^2,
\end{equation}
where  $\delta$
   is a constant defined as $
\delta  = 3/2\left( {1 - \beta } \right).$
  We repeat our procedure. However, since now they are dealing with a two-parameter model, it is necessary to include in the consideration the derivative$\ddot H.$
   Finally, we obtain free parameters of the proposed model in terms of the cosmographic parameters $H,q,j$
   \begin{align}
\nonumber \delta=\frac{2+3q+j}{2(1+q} \,\, or\,\, \beta=\frac{1-j}{3(1+q)} \\
\label{eq7.17}\Lambda_0=H^2\frac{j-q-2 q^2}{1+q}.
 \end{align}
 Thus, having treated all the parameters of our models through the directly measurable cosmographical parameters, we can calculate the original parameters . After that, solutions of evolution  equations can be analyzed using the parameters found.
The solution of the evolution equation (\ref{eq7.13})  for the first model is given by the following time-dependence of the Hubble parameter
\begin{equation}\label{eq7.18}
H(t)=\frac{H_0 \sigma}{3 H_0 +(\sigma-3 H_0 )e^{\frac{\sigma(t-t_0)}{2}}}.
\end{equation}
Here $H_0 $
 is current value of the  Hubble parameters.
For the second model, it is also possible to solve the evolution equation (\ref{eq7.16}) and to find the time-dependence of the Hubble parameter
\begin{equation}\label{eq7.19}
H(t)=\sqrt{\frac{\Lambda _0}{2 \delta}}Tanh\Bigl[\sqrt{\frac{\delta \Lambda _0}{2}}(t-t_0)+ArcTanh \Bigl(\sqrt{{\frac{2 \delta}{\Lambda _0}}}H_0\Bigr)\Bigr].
\end{equation}

\begin{figure}[pb]
\centerline{\psfig{file=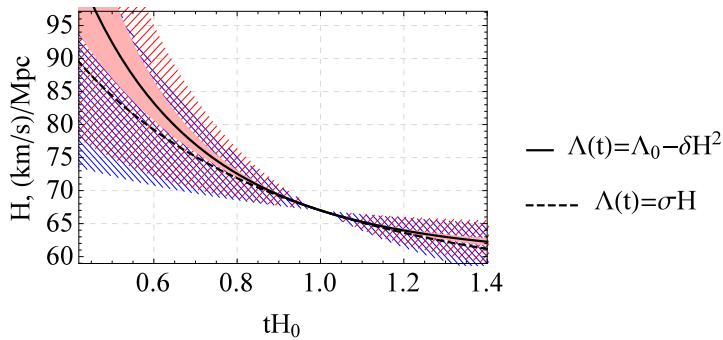,width=9.7cm}}
\vspace*{8pt}
\caption{Time-dependencies (\ref{eq7.18}) (dashed line) and (\ref{eq7.18}) (solid line) of the Hubble parameter corresponding to two different models (\ref{eq7.10}) and (\ref{eq7.15}) with various values for currently measurable $q_0$ and $j_0$ in their typical ranges. A typical range of current values of the deceleration parameter $q_0$ is shown by hatching (blue for the model (\ref{eq7.10}), and red for the model (\ref{eq7.15})). Here, the current parameter value $q_0$ varies from $-0.9$ to $-0.5$ with the central value $-0.7$. The current value for jerk parameter $j_0$ in the model (\ref{eq7.15}) varies from $0.8$ to $1.2$ (red area) with a central value of $1$.  The Hubble constant $H_0$ is set to be $67 km/s.Mpc.$ \label{plot12qj}}
\end{figure}

In the analysis of expressions (\ref{eq7.18}), and (\ref{eq7.19}), we used the recently found values of the atmospheric parameters\cite{Heneka:2018tta,Riess:1998cb,Mamon:2016dlv}:
$$q_0=-0.70\pm 0.18\,\ and\,\ j_0=0.52(+0.58-0.60).$$
The time-dependences of the Hubble parameter are depicted in Fig. \ref{plot12qj}

As we can see, both models demonstrate a close time dependence for the temporary interval $0.8H_0^{ - 1}  < t < 1.2H_0^{ - 1}. $.
However, one can notice the difference for early and late  times, which  is  quite simple to explain.
Let us now obtain for the two models under consideration the first Friedmann equation in terms of cosmographic parameters. For the first model (\ref{eq7.10}) using the derivative$\ddot H$
  we find
\begin{equation}\label{eq7.20}
2j - 2q^2  - q - 1 = 0
\end{equation}
For the second model (\ref{eq7.15}) , including into consideration $\dddot{H}$ we obtain
	\begin{equation}\label{eq7.21}
s(1 + q) + J^2  + j\left( {q^2  + 4q + 1} \right) + q\left( {2q + 1} \right) = 0
\end{equation}
 	
These expressions allow, within the framework of the considered work, to calculate any cosmographic parameter using several known lower parameters.

\subsection{Cosmographic constrain on the dark energy and dark matter coupling}
We now turn to a more general case. As we saw above interaction between the dark components in simplest linear case  phenomenologically is described by
\begin{equation}\label{eq7.22}
Q_I=3\delta H \rho_{dm},
\quad Q_{II}=3\delta H \rho_{de}, \quad Q_{III}=3\delta H (\rho_{dm}+\rho_{de}).
\end{equation}
Let us Universe filled with two components labeled 1 and 2 with the equations of state  and the interaction function respectively\cite{Bolotin:2015fea}
\begin{equation}\label{eq7.23}
p_1=w_1\rho_1,\quad p_2=w_2\rho_2,\quad Q=3 \delta \,H\,\rho_1
.
\end{equation}
In particular, the case $Q_I $
  corresponds to $w_1  = 0,\quad w_2  = w$
 and the case $Q_{II} $
 is obtained with $w_1  = w,\quad w_2  = 0$
 and  $
\delta  \to  - \delta.
$
\begin{figure}[pb]
\centerline{\includegraphics[width=10.0cm]{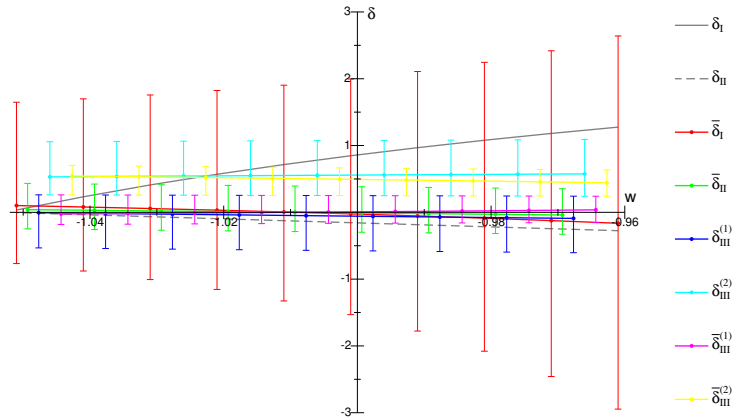}}
\caption{\label{fig1} The coupling constant $\delta$  as a function of the dark energy parameter $w$ for different interaction models. }
\end{figure}
Omitting cumbersome calculations, we present only the final expressions for the constant interaction in the flat $\delta$
and $\bar \delta$  in the general case,
\begin{align}
\nonumber \delta_I&=2+\frac{s+q\, j+3(w+1)j}{j-(3\,w+2)q} ;\\
\nonumber \delta_{II}&=2+3 \, w+\frac{s+(q+3)j}{j-2 \,q}; \\
\nonumber\bar{ \delta}_I&=\frac{1}{3}+\frac{2}{3}\frac{j-q(3\,w+2)}{3\,w+1-2\,q} ;\\
\nonumber \bar{\delta}_{II}&=\frac{1}{3}+ w+\frac{2}{3}\frac{j-2\,q}{1-2\, q};\\
\nonumber \delta_{III}^{(1,2)}&=-\frac{w}{4}\left[1 \mp \sqrt{1-\frac{8}{9 q w^2 }[s-2 q(2+3 w)+j(5+q+3 w)]}\right]; \\
\label{eq7.24} \bar{\delta}_{III}^{(1,2)}&=-\frac{w}{4}\left[1 \mp \sqrt{1-\frac{8}{9 q w^2 }[1+2j+3 w-6 q(1+w)]}\right].
 \end{align}
Fig. \ref{fig1} represents the coupling constant  in the models (2) as a function of the dark energy
 parameter in the range $−1.051 < w < −0.961$ which corresponds to the most recent observations.
 Careful analysis of the plots  reveals a remarkable feature: the functions $\delta (w)$
  corresponding to different models (I,II,III) have exact common roots, namely:
\begin{align}
\nonumber\delta_I(w_0)=\delta_{II}(w_0)=\delta_{III}^{(1)}(w_0)&=0,\ w_{01}=-\frac{s+(q+5)j-4 q}{3(j-2 q}\approx-1.052;\\
\label{eq7.25}\bar{\delta}_I(\bar{w}_{02})=\bar{\delta}_{II}(\bar{w}_{02})=\bar{\delta}_{III}^{(1)}(\bar{w}_0)&=0,\ w_{02}=-\frac{1+2 j-6 q}{3(j-2 q)}\approx-1.009.
\end{align}
We proposed a novel approach to obtain limitations on the dark energy and dark matter coupling constant. The suggested approach allowed us to express the coupling constant in terms of the cosmographic parameters.  As an example we considered three cosmological models with linear type of interaction between the dark components. 
\section{Summary} We have shown that cosmography is an effective and universal method for analyzing cosmological models. A huge number of cosmological models used to describe the evolution of the Universe gave rise to an even larger number of “independent” parameters. These parameters are usually difficult to relate to the directly observable quantities. The considered approach allows us to express the parameters of any model that satisfies the cosmological principle through a limited number of cosmographic parameters.
The cosmographic approach to finding the parameters of cosmological models has many advantages. Let's briefly dwell on them.\\
1. Universality: the method is applicable to any braid model that satisfies the cosmological principle. The procedure can be generalized to the case of models with interaction between components .\\
2. Reliability: all the obtained results are accurate, since they follow from identical transformations.\\
3. The simplicity of the procedure.\\
4. Parameters of different models are expressed through a universal set of cosmological parameters. There is no need to introduce additional parameters.\\
5. The method represents a simple test for analyzing the compatibility of different models. Since the cosmological parameters are universal, the models are compatible only in the case of a non-zero domain of intersection of their parameter space.\\
Inclusion of the higher order derivatives of the scale factor, on the one hand, reflects the continuous progress of the observational cosmology, and, on the other, is dictated by the need to describe the increasingly complex effects used to obtain the precise observational data.



\section*{Acknowledgments}
This work is supported by SFFR, Ukraine, Project No. 32367.
\bibliographystyle{ws-ijmpd}

\end{document}